\documentclass[review,12pt]{elsarticle}

\usepackage{amssymb}
\usepackage{amsthm}
\usepackage{amsmath,color}
\usepackage{mathrsfs}
\usepackage{graphicx}
\usepackage{epstopdf}
\usepackage{float}
\usepackage{caption}
\usepackage{subcaption}
\usepackage{bm}
\usepackage{bbm}
\usepackage{mathrsfs}
\usepackage{cleveref}
\usepackage{soul}
\usepackage{mnsymbol} 
\biboptions{sort&compress}
\usepackage{color,soul} 
\usepackage{color} 

\journal{.}

\makeatletter
\def\@author#1{\g@addto@macro\elsauthors{\normalsize%
    \def\baselinestretch{1}%
    \upshape\authorsep#1\unskip\textsuperscript{%
      \ifx\@fnmark\@empty\else\unskip\sep\@fnmark\let\sep=,\fi
      \ifx\@corref\@empty\else\unskip\sep\@corref\let\sep=,\fi
      }%
    \def\authorsep{\unskip,\space}%
    \global\let\@fnmark\@empty
    \global\let\@corref\@empty  
    \global\let\sep\@empty}%
    \@eadauthor={#1}
}
\makeatother

\begin{document}

\begin{frontmatter}



\title{Fracture in distortion gradient plasticity}


\author{Sandra Fuentes-Alonso\fnref{Uniovi}}

\author{Emilio Mart\'{\i}nez-Pa\~neda\corref{cor1}\fnref{IC,Cam}}
\ead{e.martinez-paneda@imperial.ac.uk}

\address[Uniovi]{Department of Construction and Manufacturing Engineering, University of Oviedo, Gij\'on 33203, Spain}

\address[IC]{Department of Civil and Environmental Engineering, Imperial College London, London SW7 2AZ, UK}

\address[Cam]{Department of Engineering, Cambridge University, CB2 1PZ Cambridge, UK}

\cortext[cor1]{Corresponding author.}

\begin{abstract}
Due to its superior modelling capabilities, there is an increasing interest in \emph{distortion} gradient plasticity theory, where the role of the plastic spin is accounted for in the free energy and the dissipation. In this work, distortion gradient plasticity is used to gain insight into material deformation ahead of a crack tip. This also constitutes the first fracture mechanics analysis of gradient plasticity theories adopting Nye's tensor as primal kinematic variable. First, the asymptotic nature of crack tip fields is analytically investigated. A generalised $J$-integral is defined and employed to determine the power of the singularity. We show that an inner elastic region exists, adjacent to the crack tip, where elastic strains dominate plastic strains and Cauchy stresses follow the linear elastic $r^{-1/2}$ stress singularity. This finding is verified by detailed finite element analyses using a new numerical framework, which builds upon a viscoplastic constitutive law that enables capturing both rate-dependent and rate-independent behaviour in a computationally efficient manner. Numerical analysis is used to gain further insight into the stress elevation predicted by distortion gradient plasticity, relative to conventional $J_2$ plasticity, and the influence of the plastic spin under both mode I and mixed-mode fracture conditions. It is found that Nye's tensor contributions have a weaker effect in elevating the stresses in the plastic region, while predicting the same asymptotic behaviour as constitutive choices based on the plastic strain gradient tensor. A minor sensitivity to $\chi$, the parameter governing the dissipation due to the plastic spin, is observed. Finally, distortion gradient plasticity and suitable higher order boundary conditions are used to appropriately model the phenomenon of brittle failure along elastic-plastic material interfaces. We reproduce paradigmatic experiments on niobium-sapphire interfaces and show that the combination of strain gradient hardening and dislocation blockage leads to interface crack tip stresses that are larger than the theoretical lattice strength, rationalising cleavage in the presence of plasticity at bi-material interfaces. 
\end{abstract}


\end{frontmatter}


\section{Introduction}
\label{Introduction}

In recent years, there has been an increasing interest in characterising the behaviour
of metals at the micrometer scale. Examples are found in microelectromechanical systems (MEMS), microelectronic components, and thin film applications. A wide array of micron scale experiments have revealed that metals display pronounced size effects when deformed non-uniformly into the plastic range (see Ref. \cite{Voyiadjis2019} for a review). Notable pioneering examples are the wire torsion experiments by Fleck and co-workers \cite{Fleck1994}, the nanoindentation measurements by Nix and Gao \cite{Nix1998}, and the bending of foils by St\"{o}lken and Evans \cite{Stolken1998}. Fleck \textit{et al.} \cite{Fleck1994} tested very thin copper wires (with radius varying from 6 to 85 $\mu$m) under both uniaxial tension and torsion. Results revealed only a minor influence of specimen size on tensile behaviour but a systematic increase in torsional strengthening with decreasing wire diameter. Nix and Gao \cite{Nix1998} found a linear relation between the indentation depth and the hardness of single crystal and cold worked polycrystalline copper. This size effect becomes negligible as the indentation depth is increased beyond a characteristic length on the order of micrometers. St\"{o}lken and Evans \cite{Stolken1998} conducted micro-bending tests on nickel foils of different thicknesses, showing that thinner specimens are stronger and strain harden more than thicker ones. Uniaxial tension tests were also conducted and, as in the work by Fleck \textit{et al.} \cite{Fleck1994}, almost no influence of specimen size is observed. Thus, the \emph{smaller is harder} or \emph{smaller is stronger} trends observed in the aforementioned micron scale experiments are intrinsically associated with the presence of strain gradients. In terms of the underpinning dislocation phenomena, work hardening is controlled by the total density of dislocations, part of which is related to the gradients of plastic strain. Thereby, dislocation storage governing material hardening is due to: (i) dislocations that trap each other in a random way and (ii) dislocations required for compatible deformation of various parts of the crystal \cite{Ashby1970}. The latter are referred to as Geometrically Necessary Dislocations (GNDs) while the former are named Statistically Stored Dislocations (SSDs). GNDs do not contribute to plastic strain but to material work hardening by acting as obstacles to the motion of SSDs. This extra storage of dislocations associated with gradients of plastic strain will manifest its influence when the characteristic length of deformation becomes sufficiently small.\\

Experimental evidence of strain gradient hardening has been accompanied by a vast literature on the development of enriched isotropic plasticity models, so-called strain gradient plasticity theories (see, e.g. \cite{Berdichevsky1967a,Dillon1970,Aifantis1992,Fleck1993,Fleck2001,Gudmundson2004,
Gurtin2005,Idiart2009} and references therein). Consistent with experimental observations, theoretical models are cast in a form where the plastic work depends on both strains and strain gradients; introducing a material length scale $\ell$, and reducing to conventional plasticity when the length scales of the imposed deformation gradients are large compared to $\ell$. Recent theoretical developments have been aimed at capturing a wide range of experimental observations. Thus, modern strain gradient plasticity formulations consider both dissipative (or unrecoverable) and energetic (or recoverable) gradient contributions, to capture both the strengthening and hardening effects observed \cite{Gudmundson2004,Gurtin2005,Fleck2009}. In addition, the form of the free energy has received particular interest \cite{Ohno2007,Garroni2010,Wulfinghoff2015a,Lancioni2015,Bardella2015,Panteghini2019}. Gradient effects are accounted for by means of an additional contribution to the Helmholtz free energy, the so-called defect energy. Choices include the use of plastic strains or Nye's tensor as primal higher order kinematic variables, considering one or more invariants of the primal variable, and exploring less-than-quadratic forms of the defect energy. Also, increased attention has been focused on the need to account for the plastic spin in recent years, as originally proposed by Gurtin \cite{Gurtin2004}, to properly describe plastic flow incompatibility and the associated dislocation densities. This class of gradient plasticity models is referred to as \emph{distortion} gradient plasticity, as it builds upon the incompatibility of the plastic part of the displacement gradient and the macroscopic characterisation of the Burgers vector \cite{Burgers1939,Nye1953,Fleck1997,Gurtin2004} to rigorously define Nye's tensor as: 
\begin{equation}
\alpha_{ij}=\epsilon_{jkl} \gamma_{il,k}^p \,\,\,\,\,\,\,\, \left( \boldsymbol{\alpha}=\textnormal{curl} \, \boldsymbol{\gamma}^p \right)
\end{equation} 

\noindent where $\gamma_{ij}^p$ is the plastic distortion - the plastic part of the displacement gradient. The increasing popularity of distortion gradient plasticity lies on its superior modelling capabilities. As shown by Bardella and co-workers \cite{Bardella2008,Bardella2009,Bardella2010} and Poh and Peerlings \cite{Poh2016}, the contribution of the non-symmetric plastic part of the displacement gradient plays a fundamental role in capturing essential features of crystal plasticity. Moreover, Poh and Peerlings \cite{Poh2016} showed that the localization phenomenon that takes place in Bittencourt \textit{et al.} \cite{Bittencourt2003} composite unit cell benchmark problem can only be reproduced by distortion gradient plasticity. Other recent works involve the development of new homogenization formulations \cite{Poh2013,Poh2016a} and finite element schemes \cite{IJSS2016,Panteghini2016,Panteghini2018}. However, the implications of distortion gradient plasticity on crack tip mechanics remain to be addressed.\\

Strain gradient effects are typically characterised \textit{via} micro-scale experiments but are present in any boundary value problem where the strain varies over microns. This is the case of fracture mechanics problems where, independently of the size of the cracked sample, the plastic zone adjacent to the crack tip is typically small and contains large spatial gradients of plastic deformation. The analysis of stationary cracks with strain gradient plasticity reveals that local strain gradient hardening elevates crack tip stresses far beyond conventional plasticity predictions \cite{Wei1997,Komaragiri2008,IJSS2015,IJP2016}. The stress elevation predicted by strain gradient plasticity provides a rationale for brittle fracture in the presence of plasticity \cite{Wei1997,JMPS2019} and has important implications for a number of structural integrity problems, such as hydrogen embrittlement \cite{AM2016,JMPS2020}, fatigue damage \cite{Brinckmann2008,Pribe2019}, and low-temperature cleavage \cite{Qian2011,EJMAS2019b}. In some of these applications, quantitative differences are of utmost importance. For example in hydrogen embrittlement, where the crack tip hydrogen content has an exponential dependence on the hydrostatic stress \cite{IJHE2016}. However, fracture studies have focused on a few gradient plasticity models and the impact of recent theoretical developments is yet to be investigated.\\

In this work, distortion gradient plasticity is used for the first time to model crack tip behaviour. Also, the role of Nye's tensor on fracture mechanics is first elucidated. 
We combine analytical insight into the asymptotic stress field with detailed finite element analysis of crack tip fields under mode I and mixed mode fracture conditions, revealing some remarkable results. The remaining of the manuscript starts by introducing the flow theory of distortion gradient plasticity (Section \ref{Sec:DGPtheory}). This is followed by an analytical investigation of the nature of the asymptotic crack tip solution in Section \ref{Sec:AResults}. Our analysis reveals the existence of an \emph{elastic} region close to the crack tip, reminiscent of a dislocation-free zone. This feature is confirmed by numerical computations and extends the recent findings by Mart\'{\i}nez-Pa\~neda and Fleck \cite{EJMAS2019} to distortion gradient plasticity and strain gradient theories based on Nye's dislocation density tensor. In Section \ref{Sec:FEformulation}, the finite element framework is described, including the development of a new viscoplastic potential. The numerical model is then used in Section \ref{Sec:FEMresults} to characterize the influence of the plastic spin and material parameters on crack tip fields. It is shown that the use of a defect energy based on Nye's tensor leads to a much weaker stress elevation relative to defect energies employing the plastic strain tensor as primal variable. Finally, insight is shed into the conundrum of atomic decohesion at metal-ceramic interfaces by modelling the classic experiments by Elssner \textit{et al.} \cite{Elssner1994} and Korn \textit{et al.} \cite{Korn2002}. The manuscript ends with concluding remarks in Section \ref{Sec:ConcludingRemarks}. 

\section{The flow theory of distortion plasticity plasticity}
\label{Sec:DGPtheory}

The equations of this section refer to the mechanical response of a body occupying a space region $\Omega$ with an external surface $S$ of outward normal $n_i$. More details about the higher order theory of distortion gradient plasticity can be found in Ref. \cite{Gurtin2004}.

\subsection{Variational principles and balance equations}

Within a small strain formulation, the displacement gradient $u_{i,j}$ can be decomposed into its elastic and plastic parts:
\begin{equation}
u_{i,j} = \gamma_{ij}^e + \gamma_{ij}^p
\end{equation}

Where $\gamma_{ij}^p$, the plastic distortion, which characterises the evolution of dislocations and other defects through the crystal structure, may in turn be decomposed into its symmetric and skew parts: 
\begin{equation}
\gamma_{ij}^p = \varepsilon_{ij}^p + \vartheta_{ij}^p
\end{equation}

Unlike the plastic strain field $\varepsilon_{ij}^p$, the plastic rotation $\vartheta_{ij}^p$ is essentially irrelevant in a conventional theory. However, as pointed out by Gurtin \cite{Gurtin2004}, phenomenological models involving Nye's dislocation density tensor $\alpha_{ij}$ as primal higher order kinematic variable,
\begin{equation}
\alpha_{ij}=\epsilon_{jkl} \gamma_{il,k}^p \,\,\,\,\,\,\,\, \left( \boldsymbol{\alpha}=\textnormal{curl} \, \boldsymbol{\gamma}^p \right)
\end{equation} 

\noindent must account for the plastic spin since the macroscopic characterisation of the Burgers vector involves both the symmetric and skew parts of the plastic distortion
\begin{equation}
\epsilon_{jkl} \gamma_{il,k}^p = \epsilon_{jkl} \varepsilon_{il,k}^p + \epsilon_{jkl} \vartheta_{il,k}^p \,\,\,\,\,\, \left( \textnormal{curl} \, \boldsymbol{\gamma}^p = \textnormal{curl} \, \boldsymbol{\varepsilon}^p + \textnormal{curl} \, \boldsymbol{\vartheta}^p \right)
\end{equation}

\noindent with $\epsilon_{jkl}$ denoting the alternating symbol. The internal virtual work reads:
\begin{equation}\label{eq:iVW}
 \delta W_i = \int_\Omega \Big( \sigma_{ij} \delta \varepsilon_{ij}^e + \zeta_{ij} \delta \alpha_{ij} + S_{ij} \delta \gamma_{ij}^p + \tau_{ijk} \delta \varepsilon_{ij,k}^p \Big) \, \textnormal{d}V 
\end{equation}

\noindent where the Cauchy stress is denoted by $\sigma_{ij}$. In addition to conventional stresses, the principle of virtual work incorporates the so-called micro-stress tensor, $S_{ij}$ (work conjugate to the plastic distortion, $\gamma_{ij}^p$), the defect stress $\zeta_{ij}$ (work conjugate to Nye's tensor $\alpha_{ij}$, the curl of the plastic distortion) and the - here, purely dissipative - higher order stress tensor, $\tau_{ijk}$ (work conjugate to the plastic strain gradients $\varepsilon_{ij,k}^p$). By taking into account that the micro-stress tensor can be decomposed into its symmetric and skew parts: $S_{ij}=q_{ij}+\omega_{ij}$, the internal virtual work statement can be expressed as: 
\begin{equation}\label{eq:iVW2}
 \delta W_i = \int_\Omega \Big( \sigma_{ij} \delta \varepsilon_{ij} +\zeta_{ij} \delta \alpha_{ij}+ \left( q_{ij} - \sigma_{ij}' \right) \delta \varepsilon_{ij}^p + \omega_{ij} \delta \vartheta_{ij}^p +  \tau_{ijk} \delta \varepsilon_{ij,k}^p \Big) \, \textnormal{d}V
\end{equation}

\noindent with the prime symbol $'$ denoting deviatoric quantities. Applying Gauss' divergence theorem to (\ref{eq:iVW2}) renders:
\begin{align}\label{eq:iVW3}
\delta W_i & = \int_S \Big(\sigma_{ij} n_j  \delta u_i + \left( \Upsilon_{ij}' +  \tau_{ijk} n_k \right)  \delta \varepsilon_{ij}^p +  \Delta_{ij}  \delta \vartheta_{ij}^p \Big) \, \textnormal{d}S\nonumber \\
& -\int_\Omega \Big(  \sigma_{ij,j} \delta u_i - \left( q_{ij} - \sigma_{ij}' - \tau_{ijk,k} + \eta_{ij}' \right) \delta \varepsilon_{ij}^p  - \left(\omega_{ij} + \varphi_{ij} \right) \delta \vartheta_{ij}^p \Big) \, \textnormal{d}V  
\end{align}

\noindent where $\eta_{ij}$ and $\varphi_{ij}$ are, respectively, the symmetric and skew-symmetric parts of the curl of the defect stress $\xi_{ij}=\epsilon_{jkl} \zeta_{il,k}=\eta_{ij}+\varphi_{ij}$; and equivalently, $\Upsilon_{ij}$ and $\Delta_{ij}$ respectively denote the symmetric and skew-symmetric parts of the cross product of the defect stress and the outward normal $\Gamma_{ij}=\epsilon_{jkl} \zeta_{il} n_k=\Upsilon_{ij}+\Delta_{ij}$. Since the volume integral in (\ref{eq:iVW3}) should vanish for arbitrary variations, three sets of equilibrium equations are readily obtained:
\begin{align}
& \sigma_{ij,j}=0 \\
& q_{ij} - \sigma'_{ij} - \tau_{ijk,k} + \eta_{ij}'=0 \label{Eq:HOeq2}\\
& \omega_{ij} + \varphi_{ij} = 0
\end{align}

\noindent Now, identifying the surface in (\ref{eq:iVW3}) as part of the external work and considering (\ref{eq:iVW2}), the Principle Virtual Work reads:
\begin{align}\label{eq:PVW}
 &\int_\Omega \Big( \sigma_{ij} \delta \varepsilon_{ij} +\zeta_{ij} \delta \alpha_{ij}+ \left( q_{ij} - \sigma_{ij}' \right) \delta \varepsilon_{ij}^p + \omega_{ij} \delta \vartheta_{ij}^p +  \tau_{ijk} \delta \varepsilon_{ij,k}^p  \Big) \, \textnormal{d}V \nonumber \\
& = \int_S \Big( T_i \delta u_i + t_{ij}^\varepsilon \delta \varepsilon^p_{ij} +  t_{ij}^\vartheta \delta \vartheta^p_{ij} \Big) \, \textnormal{d}S  
\end{align}

\noindent where $T_{i}$ are the conventional tractions, work conjugate to the displacements, while $t_{ij}^\varepsilon$ and $t_{ij}^\vartheta$ denote the higher order tractions work conjugate to plastic strains $\varepsilon^p_{ij}$ and plastic rotations $\vartheta^p_{ij}$, respectively. Accordingly, considering (\ref{eq:iVW3}), the natural boundary conditions read:
\begin{align}
& T_i=\sigma_{ij} n_j \\
& \Upsilon_{ij}' +  \tau_{ijk} n_k= t_{ij}^\varepsilon \\
& \Delta_{ij}=t_{ij}^\vartheta
\end{align}

\subsection{Energetic contributions}

In order to account for the influence of GNDs, the free energy is chosen to depend on both the elastic strain $\varepsilon_{ij}^e$ and Nye's tensor $\alpha_{ij}$:
\begin{equation}
\Psi=\frac{1}{2}C_{ijkl} \varepsilon_{ij}^e \varepsilon_{kl}^e + \Phi \left( \alpha_{ij} \right)
\end{equation}

\noindent with $C_{ijkl}$ being the elastic stiffness and $\Phi \left( \alpha_{ij} \right)$ the defect energy that accounts for the recoverable mechanisms associated with the development of GNDs. The widely used quadratic form of the defect energy is adopted
\begin{equation}\label{Eq:DefectEnGurtin}
\Phi \left( \alpha_{ij} \right)= \frac{1}{2} \mu L_E^2 \alpha_{ij} \alpha_{ij}
\end{equation}

\noindent but one should note that exploring other options may lead to further modelling capabilities \cite{Ohno2007,Garroni2010,Wulfinghoff2015a,Lancioni2015,Bardella2015,Panteghini2019}. Accordingly, the defect stress equals:
\begin{equation}\label{eq:defectstress}
\zeta_{ij}=\frac{\partial \Phi \left( \alpha_{ij} \right)}{\partial \alpha_{ij}}= \mu L_E^2 \alpha_{ij}
\end{equation}

\noindent with $\mu$ being the shear modulus and $L_E$ the energetic material length scale. 

\subsection{Dissipative contributions}

A gradient-enhanced phenomenological effective plastic flow rate is defined,
\begin{equation}\label{Eq:EpGurtin}
\dot{E}^p=\sqrt{\frac{2}{3} \dot{\varepsilon}_{ij}^p \dot{\varepsilon}_{ij}^p +\chi \dot{\vartheta}_{ij}^p \dot{\vartheta}_{ij}^p+ \frac{2}{3}L_D^2 \dot{\varepsilon}_{ij,k}^p \dot{\varepsilon}_{ij,k}^p}
\end{equation}

\noindent where $L_D$ is a dissipative length parameter and $\chi$ is the parameter governing the dissipation due to the plastic spin. Bardella \cite{Bardella2009} has analytically identified the value of $\chi$ that captures the mechanical response of a crystal subjected to multi-slip under simple shear:
\begin{equation}
\chi=\left[\frac{3}{2} + \frac{\sigma_Y}{\mu \varepsilon_Y} \left(\frac{L_D}{L_E} \right)^2 \right]^{-1}
\end{equation}

\noindent being $\sigma_0$ and $\varepsilon_0$ non-negative material parameters, which implies a value for $\chi$ bounded between $0$ and $2/3$. The flow resistance $\Sigma$, work conjugate to $\dot{E}^p$, is given by
\begin{equation}
\Sigma=\sqrt{\frac{3}{2}q_{ij}q_{ij}+\frac{1}{\chi}\omega_{ij}\omega_{ij}+\frac{3}{2L_D^2} \tau_{ijk} \tau_{ijk}}
\end{equation}

Such that the unrecoverable stresses equal
\begin{equation}\label{eq:DisspStress}
q_{ij}=\frac{2}{3} \frac{\Sigma}{\dot{E}^p} \dot{\varepsilon}_{ij}^p, \,\,\,\,\,\,\,\,\,\,\,\,\,\,\,\,
\omega_{ij} = \chi \frac{\Sigma}{\dot{E}^p} \dot{\vartheta}_{ij}^p, \,\,\,\,\,\,\,\,\,\,\,\,\,\,\,\,
\tau_{ijk}=\frac{2}{3} L_D^2 \frac{\Sigma}{\dot{E}^p} \dot{\varepsilon}_{ij,k}^p
\end{equation}

And consequently the second law of thermodynamics is fulfilled by relating finite stress measures with rates of plastic deformation, in what is referred to as a \emph{non-incremental} form:
\begin{equation}
q_{ij} \dot{\varepsilon}_{ij}^p+\omega_{ij} \dot{\vartheta}_{ij}^p + \tau_{ijk} \dot{\varepsilon}_{ij,k} \equiv \Sigma \dot{E}^p > 0
\end{equation} 

\section{Asymptotic analysis of crack tip fields}
\label{Sec:AResults}

We begin our study by conducting an asymptotic analysis of the relevant fields at the crack tip under mode I fracture conditions. Consider a crack in a 2D space, with its tip at the origin of a polar coordinate system $(r, \, \theta)$. We will assume that the plastic distortion field $\gamma^p_{ij}$ is continuous and differentiable, with an asymptotic solution that behaves as follows:
\begin{equation}\label{eq:SolutionGamma}
\gamma^p_{ij} \sim r^\beta f_{ij}(\theta)
\end{equation}

\noindent for $r \to 0$. By deriving a generalized $J$-integral for distortion gradient plasticity, the order of the singularity (index $\beta$) will be determined using energy boundness arguments and its implications for the behaviour of crack tip stresses investigated. 

\subsection{Deformation theory solid}
\label{Sec:DeformationTheory}

In a deformation theory context, in the absence of conventional and higher order tractions, the total potential energy assumes the form,
\begin{equation}
U \left( u_i, \, \varepsilon_{ij}^p, \, \vartheta_{ij}^p, \, \varepsilon_{ij,k}^p, \, \alpha_{ij} \right) = \int_V \, \left[ \Psi \left( u_i, \, \alpha_{ij} \right) + \varphi \left( \varepsilon_{ij}^p, \, \vartheta_{ij}^p, \, \varepsilon_{ij,k}^p \right) \right] \textnormal{d}V
\end{equation}

\noindent with the free energy being given by,
\begin{equation}
\Psi \left( u_i, \, \alpha_{ij} \right) = \Psi^e (u_i) + \Phi \left( \alpha_{ij} \right) = \frac{1}{2} \left( \varepsilon_{ij} - \varepsilon_{ij}^p \right) C_{ijkl} \left( \varepsilon_{kl} - \varepsilon_{kl}^p \right) + \frac{1}{2} \mu L_E^2 \alpha_{ij} \alpha_{ij}
\end{equation}

\noindent Here, $\Psi^e$ denotes the elastic free energy. Thereby, the Cauchy stresses are derived as,
\begin{equation}
\sigma_{ij}=\frac{\partial \Psi}{\partial \varepsilon_{ij}^e}=C_{ijkl}\left( \varepsilon_{kl} - \varepsilon_{kl}^p \right)
\end{equation}

\noindent and the so-called defect stress $\zeta_{ij}$ is given by (\ref{eq:defectstress}).

On the other hand, the dissipation potential $\varphi$ is given by,
\begin{equation}\label{eq:disspotential}
\varphi \left( \varepsilon_{ij}^p, \, \vartheta_{ij}^p, \, \varepsilon_{ij,k}^p \right) = \frac{\sigma_Y \varepsilon_Y}{N+1} \left( \frac{E^p \left( \varepsilon^p_{ij}, \vartheta^p_{ij}, \varepsilon^p_{ij,k}\right)}{\varepsilon_Y} \right)^{N+1}
\end{equation}

\noindent where $\sigma_Y$ is the yield stress, $\varepsilon_Y$ is the yield strain, and
\begin{equation}\label{eq:EpDefTheory}
\left( E^p \right)^2 = \frac{2}{3} \varepsilon_{ij}^p \varepsilon_{ij}^p + \chi \vartheta_{ij}^p \vartheta_{ij}^p + \frac{2}{3} L_D^2 \varepsilon_{ij,k}^p \varepsilon_{ij,k}^p
\end{equation}

The choice (\ref{eq:disspotential}) implies that a homogeneous hardening law relates $E^p$ with its work conjugate, the effective stress $\Sigma$,
\begin{equation}
\Sigma = \sigma_Y \left( \frac{E^p}{\varepsilon_Y} \right)^N = \Sigma_0 \left( E^p \right)^N
\end{equation}

\noindent where $0 \leq N \leq 1$ is the strain hardening exponent. Hence, the dissipation potential reads, 
\begin{equation}
\varphi \left( \varepsilon_{ij}^p, \, \vartheta_{ij}^p, \, \varepsilon_{ij,k}^p \right) = \frac{\Sigma E^p}{N+1}
\end{equation}

\noindent Accordingly, the constitutive relations for the deformation theory solid can be readily derived as
\begin{equation}
q_{ij} = \frac{\partial \varphi}{\partial \varepsilon_{ij}^p} =  \sigma_Y \left( \frac{E^p}{\varepsilon_Y} \right)^N \frac{2}{3} \frac{\varepsilon_{ij}^p}{E^p} = \frac{2}{3} \frac{\Sigma}{E^p} \varepsilon_{ij}^p
\end{equation}

\begin{equation}
\omega_{ij} = \frac{\partial \varphi}{\partial \vartheta_{ij}^p} =   \sigma_Y \left( \frac{E^p}{\varepsilon_Y} \right)^N \chi \frac{\vartheta_{ij}^p}{E^p} = \chi \frac{\Sigma}{E^p} \vartheta_{ij}^p
\end{equation}

\begin{equation}
\tau_{ijk} = \frac{\partial \varphi}{\partial \varepsilon_{ij,k}^p} = \sigma_Y \left( \frac{E^p}{\varepsilon_Y} \right)^N L_D^2 \frac{2}{3} \frac{\varepsilon_{ij,k}^p}{E^p} = \frac{2}{3} L_D^2 \frac{\Sigma}{E^p} \varepsilon_{ij,k}^p
\end{equation}

\subsection{A generalized J-integral for distortion gradient plasticity}
\label{Sec:J-integral}

We proceed to define a generalized $J$-integral for distortion gradient plasticity. Consider a Cartesian coordinate system ($x,y$) with the crack tip at the origin and the crack plane along the negative $x$ axis. Defining $J$ as the energy release rate per unit crack extension and $w$ as the strain energy density of the solid, an evaluation of $J$ over a contour $\Gamma$ that encloses the crack tip gives
\begin{equation}\label{eq:Jintegral}
J = \int_\Gamma \left( w n_x - \sigma_{ij} n_j u_{i,x}  - t^\vartheta_{ij} \vartheta_{ij}^p - t_{ij}^\varepsilon \varepsilon^p_{ij} \right) \, \text{d} S
\end{equation}

The derivation and proof are straightforward and follow the works by Eshelby \cite{Eshelby1956} and Rice \cite{Rice1968a} in the context of conventional deformation solids, and the recent work by Mart\'{\i}nez-Pa\~neda and Fleck \cite{EJMAS2019} for strain gradient solids. Note that the existence of a $J$-integral implies that total strain energy density of the solid will asymptotically behave as $w\sim J/r$ so as to give a finite energy release rate $J$ at the crack tip. Following the notation of Section \ref{Sec:DeformationTheory}, this energy boundness constraint can be expressed as: 
\begin{equation}\label{eq:EnergyBoundness}
\Psi^e \left( \varepsilon_{ij}^e \right) + \Phi \left( \alpha_{ij} \right) + \varphi \left( \varepsilon_{ij}^p, \, \vartheta_{ij}^p, \, \varepsilon_{ij,k}^p \right) \sim \frac{J}{r}
\end{equation}

\noindent for $r \to 0$.

\subsection{Asymptotic crack tip fields}

We proceed to make use of the constitutive relations and the energy boundness constraint (\ref{eq:EnergyBoundness}) to obtain the singularity power index $\beta$ in (\ref{eq:SolutionGamma}). Further, we will make use of the higher order equilibrium equation (\ref{Eq:HOeq2}) to determine the singularity order of the Cauchy stress $\sigma_{ij}$. Note that (\ref{Eq:HOeq2}) involves $\tau_{ijk,k}$ and the symmetric part of the curl of the defect stress. Thus, from (\ref{eq:SolutionGamma}), the analysis requires obtaining the solution for: (i) the curl of Nye's tensor, which is obtained from the curl of the plastic distortion, (ii) the plastic strain gradients, and (iii) the Laplacian of the plastic strain. Consider an incompressible solid where the asymptotic solution for $\gamma^p_{ij}$ is given by (\ref{eq:SolutionGamma}); in a polar coordinate system, the individual components of the plastic distortion tensor read, 
\begin{equation}
\gamma_{rr}^p=-\gamma_{\theta \theta}^p \sim r^\beta f_1 \left( \theta \right) \, ; \,\,\,\,\,\,\,\,\,\,\,\,\,\,\,\, \gamma_{r \theta}^p \sim r^\beta f_2 \left( \theta \right) \, ; \,\,\,\,\,\,\,\,\,\,\,\,\,\,\,\,  \gamma_{\theta r}^p \sim r^\beta f_3 \left( \theta \right)
\end{equation}

Accordingly, the solution for the plastic strain and plastic spin fields is of the following form,
\begin{equation}
\varepsilon_{rr}^p=-\varepsilon_{\theta \theta}^p \sim r^\beta f_1 \left( \theta \right) \, ; \,\,\,\,\,\,\,\,\,\,\,\,\,\,\,\, \varepsilon_{r \theta}^p \sim r^\beta f_4 \left( \theta \right) \, ; \,\,\,\,\,\,\,\,\,\,\,\,\,\,\,\, \vartheta_{r \theta}^p \sim r^\beta f_5 \left( \theta \right) 
\end{equation} 

And the relevant components of the plastic strain gradient and the Laplacian of the plastic strain readily follow,
\begin{equation}
\varepsilon^p_{rr,r} = \frac{\partial \varepsilon^p_{rr}}{\partial r} = \beta r^{\beta -1 } f_1 \left( \theta \right) ; \,\,\,\,\, \varepsilon^p_{rr,\theta} = \frac{1}{r} \left(  \frac{\partial \varepsilon^p_{rr}}{\partial \theta} - 2  \varepsilon^p_{r \theta} \right)  = r^{\beta-1} \left( f_1' \left( \theta \right) - 2 f_4 \left( \theta \right) \right)
\end{equation}
\begin{equation}
\varepsilon^p_{r \theta, r} =  \frac{\partial \varepsilon^p_{r\theta}}{\partial r} =  \beta r^{\beta -1 } f_4 \left( \theta \right) ; \,\,\,\,\, \varepsilon^p_{r \theta, \theta} = \frac{1}{r} \left( \frac{\partial \varepsilon^p_{r \theta}}{\partial \theta}+ 2\varepsilon^p_{rr} \right) =  r^{\beta-1} \left( f_4' \left( \theta \right) + 2 f_1 \left( \theta \right) \right)
\end{equation}
\begin{align}\label{eq:eprr,kk}
\varepsilon^p_{rr,kk} &= \frac{\partial^2 \varepsilon^p_{rr}}{\partial r^2} + \frac{1}{r} \frac{\partial \varepsilon_{rr}^p}{\partial r} + \frac{1}{r^2}\frac{\partial^2 \varepsilon_{rr}^p}{\partial \theta^2} -  \frac{4}{r^2}\left( \varepsilon^p_{r r} +  \frac{\partial \varepsilon^p_{r \theta}}{\partial \theta}\right) \\ \nonumber 
&= r^{\beta-2} \left[ f_1 \left( \theta \right) \left( \beta^2 -4 \right) +  f_1'' \left( \theta \right) + 4 f'_4 \left( \theta \right) \right]
\end{align}
\begin{align}\label{eq:eprtheta,kk}
\varepsilon^p_{r \theta,kk} & = \frac{\partial^2 \varepsilon^p_{r\theta}}{\partial r^2} +  \frac{1}{r} \frac{\partial \varepsilon_{r\theta}^p}{\partial r} + \frac{1}{r^2} \frac{\partial^2 \varepsilon^p_{r \theta}}{\partial \theta^2} + \frac{4}{r^2} \left(\frac{\partial \varepsilon^p_{rr}}{\partial \theta} -  \varepsilon^p_{r \theta} \right) \\ \nonumber 
&= r^{\beta-2} \left[ f_4 \left( \theta \right) \left( \beta^2 -4 \right) +  f_4'' \left( \theta \right) + 4 f'_1 \left( \theta \right) \right]
\end{align}

Furthermore, the relevant components of Nye's tensor in polar coordinates read
\begin{align}\label{eq:AsympNye1}
\alpha_{r z} &=\left( \text{curl} \, \gamma_{ij}^p \right)_{rz} = \frac{\partial \gamma^p_{r\theta}}{\partial r} - \frac{1}{r} \left[ \frac{\partial \gamma^p_{rr}}{\partial \theta} - \left(  \gamma^p_{r \theta}+ \gamma^p_{\theta r}\right) \right]  \\ \nonumber
& = r^{\beta-1} \left[ f_2 \left( \theta \right) \left( \beta + 1 \right) - f_1' \left( \theta \right) + f_3 \left( \theta \right) \right] 
\end{align}
\begin{align}\label{eq:AsympNye2}
\alpha_{\theta z} &= \left( \text{curl} \, \gamma_{ij}^p \right)_{\theta z} =\frac{\partial \gamma^p_{\theta \theta}}{\partial r} - \frac{1}{r} \left( \frac{\partial \gamma^p_{r \theta}}{\partial \theta}+2 \gamma^p_{rr} \right) \\ \nonumber
& = r^{\beta-1} \left[ f_1 \left( \theta \right) \left( -2 - \beta \right) - f_2' \left( \theta \right) \right]
\end{align}

Finally, the components related to the curl of the defect stress, $\xi_{ij}=\text{curl}  \left( \mu L_E^2 \, \text{curl} \, \gamma_{ij}^p \right)=\eta_{ij}+\varphi_{ij}$, are obtained as

\begin{align}\label{eq:AsymCurlDefect1}
\xi_{rr} &= \frac{1}{r} \left\{ \frac{\partial \left[\frac{\partial \gamma^p_{r\theta}}{\partial r} - \frac{1}{r} \left( \frac{\partial \gamma^p_{rr}}{\partial \theta} -  \gamma^p_{r \theta} - \gamma^p_{\theta r}\right) \right]}{\partial \theta} - \frac{\partial \gamma^p_{\theta \theta}}{\partial r} + \frac{1}{r} \left( \frac{\partial \gamma^p_{r \theta}}{\partial \theta}+ 2\gamma^p_{rr} \right) \right\}  \\ \nonumber 
& = r^{\beta-2} \left[ f_1 \left( \theta \right) \left( 2 + \beta \right) + f_2' \left( \theta \right) \left( 2 + \beta \right) - f_1'' \left( \theta \right) + f_3' \left( \theta \right) \right]
\end{align}
\begin{equation}
\xi_{r \theta} =  - \frac{\partial \left[\frac{\partial \gamma^p_{r\theta}}{\partial r} - \frac{1}{r} \left( \frac{\partial \gamma^p_{rr}}{\partial \theta} -  \gamma^p_{r \theta}- \gamma^p_{\theta r} \right) \right]}{\partial r}= r^{\beta-2} \left( 1- \beta \right) \left[ f_2 \left( \theta \right) \left( 1 + \beta \right) - f_1' \left( \theta \right) +f_3 \left( \theta \right) \right]
\end{equation}
\begin{align}
\xi_{\theta r} &= \frac{1}{r} \left\{ \frac{\partial \left[ \frac{\partial \gamma^p_{\theta \theta}}{\partial r} - \frac{1}{r} \left( \frac{\partial \gamma^p_{r \theta}}{\partial \theta}+ 2\gamma^p_{rr} \right) \right]}{\partial \theta} +  \frac{\partial \gamma^p_{r\theta}}{\partial r} - \frac{1}{r} \left( \frac{\partial \gamma^p_{rr}}{\partial \theta} -  \gamma^p_{r \theta} -  \gamma^p_{\theta r}\right) \right\}  \\ \nonumber 
& =  r^{\beta-2} \left[ f_1' \left( \theta \right) \left( -3 - \beta \right) + f_2 \left( \theta \right) \left( \beta + 1 \right) - f_2'' \left( \theta \right) + f_3 \left( \theta \right) \right]
\end{align}
\begin{equation}\label{eq:AsymCurlDefect2}
\xi_{\theta \theta} = - \frac{\partial \left[ \frac{\partial \gamma^p_{\theta \theta}}{\partial r} - \frac{1}{r} \left( \frac{\partial \gamma^p_{r \theta}}{\partial \theta}+ 2\gamma^p_{rr}  \right) \right]}{\partial r} = r^{\beta-2} \left( 1- \beta \right) \left[ f_1 \left( \theta \right) \left(- 2 - \beta \right) - f_2' \left( \theta \right) \right]
\end{equation}

We proceed to determine the index of the singularity, $\beta$, neglecting the angular functions. Thus, we estimate the singular order of the elastic strain energy density $\Psi^e$, the defect energy $\Phi$ and the dissipation potential $\varphi$ and take into consideration that the energy released at the crack tip must be finite, see Section \ref{Sec:J-integral}. We consider the general case, ($L_D\neq 0$, $L_E \neq 0$), and particularize later.\\

The gradient term is more singular and dominates the asymptotic behaviour of the generalized plastic strain $E^p$, see (\ref{eq:EpDefTheory}); accordingly,
\begin{equation}
E^p \sim L_D \varepsilon_{ijk}^p \sim r^{\beta-1}
\end{equation} 

And its work conjugate stress reads,
\begin{equation}
\Sigma = \Sigma_0 \left( E^p \right)^N \sim r^{N \left( \beta - 1 \right)}
\end{equation}

Consequently, the asymptotic behaviour of the dissipative stresses associated with the primal kinematic variable $\varepsilon_{ij}^p$ is given by,
\begin{equation}
q_{ij} = \frac{2}{3} \frac{\Sigma}{E^p} \varepsilon_{ij}^p \sim r^{N + \beta} \, ; \,\,\,\,\,\,\,\,\,\,\, \tau_{ijk} = \frac{2}{3} L_D^2 \frac{\Sigma}{E^p} \varepsilon_{ij,k}^p \sim r^{N +\beta -1}
\end{equation}

Now consider (\ref{eq:AsympNye1})-(\ref{eq:AsympNye2}); the asymptotic behaviour of the energetic defect stress reads,
\begin{equation}
\zeta_{ij} = \mu L_E^2 \alpha_{ij}=\mu L_E^2  \epsilon_{jkl} \gamma_{il,k}^p \sim r^{\beta -1 }
\end{equation}

Finally, the asymptotic behaviour of the Cauchy stress is obtained from the higher order equilibrium equation (\ref{Eq:HOeq2}), which involves the Laplacian of the plastic strains \textit{via} the dissipative term $\tau_{ijk,k}$ and the curl of the defect stress \textit{via} the energetic term $\eta_{ij}'$. Both terms are more singular than $q_{ij}$ and have in fact the same singularity order: $r^{\beta-2}$ - see (\ref{eq:eprr,kk})-(\ref{eq:eprtheta,kk}) and (\ref{eq:AsymCurlDefect1})-(\ref{eq:AsymCurlDefect2}). Hence,
\begin{equation}
\sigma_{ij}' = q_{ij} - \tau_{ijk,k} + \eta_{ij}' \sim r^{\beta-2}
\end{equation}

In other words, the singularity exhibited by the crack tip stresses will be the same if only energetic higher order terms are present ($L_E \neq 0$, $L_D=0$) and if only dissipative higher order terms are present ($L_D \neq 0$, $L_E=0$). The use of Nye's tensor as primal higher order kinematic variable leads to identical asymptotic crack tip behaviour relative to the choice of a defect energy with the plastic strain tensor as primal variable. In all cases a quadratic form of the defect energy is assumed; interestingly, less-than-quadratic defect energies will have important implications in fracture problems: changing the nature of the stress singularity (if $L_D=0$) or making energetic contributions negligible relative to their dissipative counterparts (if $L_D \neq 0$).\\

Consider now the relevant energy quantities. The elastic strains will have the same asymptotic behaviour as the Cauchy stresses, and consequently:
\begin{equation}
\Psi^e = \frac{1}{2} C_{ijkl} \varepsilon_{ij}^e \varepsilon_{kl}^e \sim r^{2 (\beta -2)}
\end{equation}

While the defect energy and dissipation potential vary as,
\begin{equation}
\Phi = \frac{1}{2} \mu L_E^2 \alpha_{ij} \alpha_{ij} \sim r^{2 (\beta -1)}
\end{equation}
\begin{equation}
\varphi = \frac{\Sigma E^p}{N+1} \sim r^{(N+1) (\beta - 1)}
\end{equation}

Therefore, $\Psi^e$ is the most singular contribution and will dominate the energy released in the vicinity of the crack tip - see (\ref{eq:EnergyBoundness}). Since the total strain energy density must scale as $\sim J/r$ to give a finite energy release rate at the crack tip, we conclude that the singularity index $\beta$ must be equal to $3/2$. The implications of this finding are the following:
\begin{itemize}
  \item The elastic energy dominates as $r \to 0$ and the plastic field is not sufficiently singular to give any contribution to the energy release rate. If the plastic energy terms, $\Phi$ or $\varphi$, were to behave asymptotically as $J/r$, $\beta$ would be equal to $1$ (for $N=0$) or smaller (if $N>0$) and the energy release rate at the crack tip would be unbounded ($\Psi^e \sim w \sim r^{-2}$).
  \item Crack tip stresses follow the linear elastic $r^{-1/2}$ singularity, revealing the existence of an inner elastic $K$-field that is reminiscent of a dislocation-free zone. 
  \item The plastic strain field $\varepsilon_{ij}^p$ tends to zero as the crack is approached. Crack tip asymptotic analyses for distortion gradient plasticity, and similar classes of gradient theories, should not be built on the assumption that plastic strains dominate elastic strains, as done for conventional plasticity (HRR field \cite{Hutchinson1968,Rice1968}) and previous studies in strain gradient plasticity \cite{Xia1996,Huang1997,Chen1999}. 
\end{itemize}

We proceed to corroborate these findings with detailed finite element analysis, as well as exploring other interesting features of distortion gradient plasticity predictions in fracture problems.

\section{Numerical formulation and solution procedure}
\label{Sec:FEformulation}

The flow theory of distortion gradient plasticity, described in Section \ref{Sec:DGPtheory}, is implemented in a robust, backward Euler finite element framework. This is largely facilitated by the definition of a new viscoplastic potential, able to model both rate-dependent and rate-independent behaviour, by extending the work of Panteghini and Bardella \cite{Panteghini2018}. 

\subsection{Viscoplastic law}

Gradient plasticity theories are commonly implemented within a rate-dependent setting, taking advantage of its well-known computational capabilities
and circumventing complications associated with identifying active plastic zones in the corresponding time independent
model \cite{Nielsen2013,Nielsen2014}. In the context of rate-dependent gradient plasticity models, an effective flow resistance $\Sigma$ is defined,
\begin{equation}\label{Eq:SfVEp}
\Sigma \left(\dot{E}^p, E^p \right)=\sigma_F \left( E^p \right) V \left( \dot{E}^p \right)
\end{equation}

\noindent which is work conjugated to the gradient-enhanced effective plastic flow rate $\dot{E}^p$. Here, $\sigma_F$ is the current flow stress, which depends on the initial yield stress $\sigma_Y$ and the hardening law. Several viscoplastic laws have been proposed in the literature; the most exploited one is arguably the following, (see, e.g., \cite{Needleman1988,IJSS2016})
\begin{equation}
\mathscr{V} \left( \dot{E}^p, E^p \right)= \frac{\sigma_F \left( E^p \right) \dot{\varepsilon}_0}{m+1} \left( \frac{\dot{E}^p}{\dot{\varepsilon}_0} \right)^{m+1}
\end{equation}

\noindent so that
\begin{equation}\label{eq:ViscoConventional}
\Sigma \left( \dot{E}^p, E^p \right)= \sigma_F V (\dot{E}^p)= \sigma_F \left( E^p \right)  \left( \frac{\dot{E}^p}{\dot{\varepsilon}_0} \right)^m
\end{equation}

\noindent with $m$ being the material rate sensitivity exponent, $\dot{\varepsilon}_0$ the reference strain rate and $V (\dot{E}^p)$ the viscoplastic function. However, under this choice the initial tangent is infinite and the derivative $\partial \Sigma / \partial \dot{E}^p$ tends to infinity if $\dot{E}^p \to 0$, making the finite element system ill-conditioned for small values of $\dot{E}^p$. To overcome these numerical issues, Panteghini and Bardella \cite{Panteghini2016} proposed the following viscoplastic function,
\begin{equation}
V (\dot{E}^p) =   \begin{cases} 
   \frac{\dot{E}^p}{2 \dot{\varepsilon}_0} & \text{if } \dot{E}^p / \dot{\varepsilon}_0 \leq 1 \\
   1 - \frac{\dot{\varepsilon}_0}{2 \dot{E}^p}     & \text{if } \dot{E}^p /\dot{\varepsilon}_0 > 1
  \end{cases}
\end{equation}

\begin{figure}[H]
\centering
\includegraphics[scale=1]{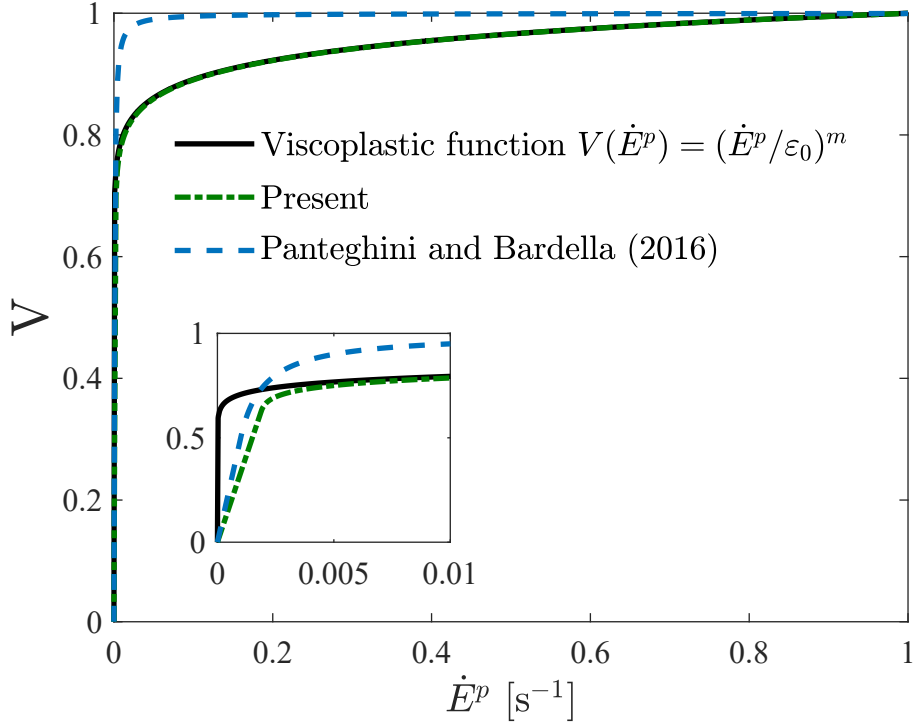}
\caption{Comparison between the classic viscoplastic power law, the viscoplastic function presented and the one proposed by Panteghini and Bardella \cite{Panteghini2016}. In the function by Panteghini and Bardella \cite{Panteghini2016} the reference strain rate $\dot{\varepsilon}_0$ equals $10^3$ s$^{-1}$, while in the other two cases a rate sensitivity exponent of $m=0.05$ and a reference strain rate $\dot{\varepsilon}_0=1$ are adopted; these choices pertain only to the present graph.}
\label{fig:Visco}
\end{figure}

In this way, the contribution of $\partial \Sigma / \partial \dot{E}^p$ will remain bounded when $\dot{E}^p \to 0$. This viscoplastic function is intended to reproduce the rate-independent limit in a robust manner, which is attained when $\dot{\varepsilon}_0 \to 0$. We extend the work by Panteghini and Bardella \cite{Panteghini2016} to develop a viscoplastic algorithm that can overcome the aforementioned numerical issues, and enables modelling both rate-dependent and rate-independent behaviour by recovering the well-known viscoplastic function $V \left( \dot{E}^p \right) = \left( \dot{E}^p / \dot{\varepsilon}_0 \right)^m$. For this purpose, a threshold effective plastic strain rate is defined $\dot{E}^p_*$ such that the viscoplastic function reads,
\begin{equation}
V (\dot{E}^p) =   \begin{cases} 
   \frac{\dot{E}^p}{\varpi \dot{\varepsilon}_0} & \text{if } \dot{E}^p m / \dot{E}^p_* \leq 1 \\
   \left( \frac{\dot{E}^p - \frac{1-m}{m} \dot{E}^p_*}{\dot{\varepsilon}_0} \right)^m & \text{if } \dot{E}^p m / \dot{E}^p_* > 1
  \end{cases}
\end{equation}

\noindent where $\varpi$ is a small positive constant ($\varpi << 1$). A smooth transition is obtained by computing the critical $\dot{E}^p_*$ from the relation between the derivatives,
\begin{equation}
\dot{E}^p_*=\dot{\varepsilon}_0 \left( \frac{1}{\varpi m} \right)^{1/(m-1)}
\end{equation}

\noindent and by offsetting the curve a distance $\dot{E}^p_*(1-m)/m$. This distance corresponds to the intersection between the abscissa axis and the tangent line at the critical point. In this way, we are able to reproduce a mechanical response that accurately follows the classic viscoplastic power law while providing a robust numerical framework. Representative curves for the aforementioned viscoplastic functions are shown in Fig. \ref{fig:Visco}; the regularisation proposed here approximates the classic viscoplastic function very well, enabling it to reproduce the rate sensitivity of metals, while retaining the robustness of the proposal by Panteghini and Bardella \cite{Panteghini2016}.

\subsection{Finite element discretisation}

The finite element framework takes displacements, plastic strains and plastic spin as the primary kinematic variables. Adopting symbolic and Voigt notation, the nodal variables for the displacement field $\boldsymbol{\hat{u}}$, the plastic strains $\boldsymbol{\hat{\varepsilon}}^p$, and the plastic spin $\boldsymbol{\hat{\vartheta}}^p$ are interpolated as,
\begin{equation}
\boldsymbol{u}=\sum_{n=1}^{k} \boldsymbol{N}^{\bm{u}}_n \boldsymbol{\hat{u}}_n, \,\,\,\,\,\,\,\,\,\,\,\,\,\,\,\,\, \boldsymbol{\varepsilon}^p=\sum_{n=1}^{k} \boldsymbol{N}_n^{\bm{\varepsilon}^p} \boldsymbol{\hat{\varepsilon}}^p_n, \,\,\,\,\,\,\,\,\,\,\,\,\,\,\,\,\, \boldsymbol{\vartheta}^p=\sum_{n=1}^{k} \bm{N}_n^{\bm{\vartheta}^p} \boldsymbol{\hat{\vartheta}}^p_n
\end{equation}

\noindent Here, $N_n$ denotes the shape function associated with node $n$, for a total number of nodes $k$. Similarly, the related gradient and curl-based quantities are discretised as
\begin{equation}
\bm{\varepsilon} = \sum_{n=1}^{k} \boldsymbol{B}_n^{\bm{u}} \boldsymbol{\hat{u}}_n, \,\,\,\,\,\,\,\,\,\,\,\,\, \nabla \bm{\varepsilon}^p = \sum_{n=1}^{k} \boldsymbol{B}_n^{\bm{\varepsilon}^p} \boldsymbol{\hat{\varepsilon}}_n^p, \,\,\,\,\,\,\,\,\,\,\,\,\ \bm{\alpha} = \sum_{n=1}^{k} \left( \boldsymbol{M}_n^{\bm{\varepsilon}^p} \boldsymbol{\hat{\varepsilon}}_n^p + \boldsymbol{M}_n^{\bm{\vartheta}^p} \boldsymbol{\hat{\vartheta}}^p_n \right)
\end{equation}

\noindent with the $\bm{B}$ and $\bm{M}$ matrices given explicitly in \ref{App:FEM}.
Accordingly, one can discretise the internal virtual work (\ref{eq:iVW2}) as,
\begin{align}
\delta W_i = & \int_\Omega \Big\{ \left( \boldsymbol{B}_n^{\bm{u}} \right)^T \bm{\sigma} \delta \boldsymbol{\hat{u}}_n + \left[ \left( \bm{N}_n^{\bm{\vartheta}^p}  \right)^T \bm{\omega} + \left(\boldsymbol{M}_n^{\bm{\vartheta}^p} \right)^T \bm{\zeta} \right]  \delta \boldsymbol{\hat{\vartheta}}^p_n \\ \nonumber
& +  \left[ \left( \bm{N}_n^{\bm{\varepsilon}^p} \right)^T \left( \bm{q} - \bm{\sigma} \right) + \left( \bm{B}_n^{\bm{\varepsilon}^p} \right)^T \bm{\tau}  + \left( \boldsymbol{M}_n^{\bm{\varepsilon}^p} \right)^T \bm{\zeta} \right] \delta \boldsymbol{\hat{\varepsilon}}^p_n \Big\}  \, \textnormal{d}V
\end{align}

Differentiating the internal virtual work with respect to the variation of the nodal variables provides the residuals for each kinematic variable as:
\begin{equation}
\bm{R}_n^{\bm{u}} = \int_\Omega  \left( \boldsymbol{B}_n^{\bm{u}} \right)^T \bm{\sigma} \, \textnormal{d}V
\end{equation}
\begin{equation}
\bm{R}_n^{\bm{\varepsilon}^p} = \int_\Omega  \left[ \left( \bm{N}_n^{\bm{\varepsilon}^p} \right)^T \left( \bm{q} - \bm{\sigma} \right) + \left( \bm{B}_n^{\bm{\varepsilon}^p} \right)^T \bm{\tau}  + \left( \boldsymbol{M}_n^{\bm{\varepsilon}^p} \right)^T \bm{\zeta} \right] \, \textnormal{d}V
\end{equation}
\begin{equation}
\bm{R}_n^{\vartheta^p} = \int_\Omega \left[ \left( \bm{N}_n^{\bm{\vartheta}^p}  \right)^T \bm{\omega} + \left(\boldsymbol{M}_n^{\bm{\vartheta}^p} \right)^T \bm{\zeta} \right] \textnormal{d}V
\end{equation}

The components of the consistent tangent stiffness matrices $\bm{K}_{nm}$ are obtained by considering the constitutive relations and differentiating the residuals with respect to the incremental nodal variables. Details are given in \ref{App:FEM}. The non-linear system of equations is solved iteratively from time step $t$ to $(t + \Delta t )$ using the Newton-Raphson method,
\begin{equation}
\begin{bmatrix}
\bm{u}\\
\bm{\varepsilon}^p\\
\bm{\vartheta}^p
\end{bmatrix}_{t+\Delta t}=\begin{bmatrix}
\bm{u}\\
\bm{\varepsilon}^p\\
\bm{\vartheta}^p\\
\end{bmatrix}_{t}-\begin{bmatrix}
  \boldsymbol{K}^{u,u} & \boldsymbol{K}^{u,\varepsilon^p} & \boldsymbol{0}\\
  \boldsymbol{K}^{\varepsilon^p,u} & \boldsymbol{K}^{\varepsilon^p,\varepsilon^p} & \boldsymbol{K}^{\varepsilon^p,\vartheta^p} \\
\boldsymbol{0} & \boldsymbol{K}^{\vartheta^p, \varepsilon^p} & \boldsymbol{K}^{\vartheta^p, \vartheta^p}
 \end{bmatrix}_t^{-1} \begin{bmatrix}
\bm{R}^{\bm{u}}\\
\bm{R}^{\bm{\varepsilon}^p}\\
\bm{R}^{\vartheta}
\end{bmatrix}_{t}
\end{equation}

The present backward Euler time integration scheme follows the work by Panteghini and Bardella \cite{Panteghini2016}; see Ref. \cite{IJSS2016} for a forward Euler based implementation. The finite element framework is implemented into the commercial package ABAQUS by means of a user element (UEL) subroutine.

\section{Finite Element results}
\label{Sec:FEMresults}

The numerical model described in Section \ref{Sec:FEformulation} is employed to gain insight into the fracture behaviour of distortion gradient plasticity solids. First, the analysis will be conducted with a boundary layer configuration, under small scale yielding conditions (Section \ref{Sec:SSY}). Irrotational plastic flow and mode I fracture will be assumed first, to verify the findings of the asymptotic analysis and assess the role of Nye's tensor. Mixed mode conditions are then considered to address the role of the plastic spin. Finally, fracture along a bi-material interface is investigated by reproducing the four-point bending experiments by Korn \textit{et al.} \cite{Korn2002} with suitable higher order boundary conditions (Section \ref{Sec:Bimaterial}). 

\subsection{Small scale yielding}
\label{Sec:SSY}

\begin{figure}[H]
\centering
\includegraphics[scale=0.25]{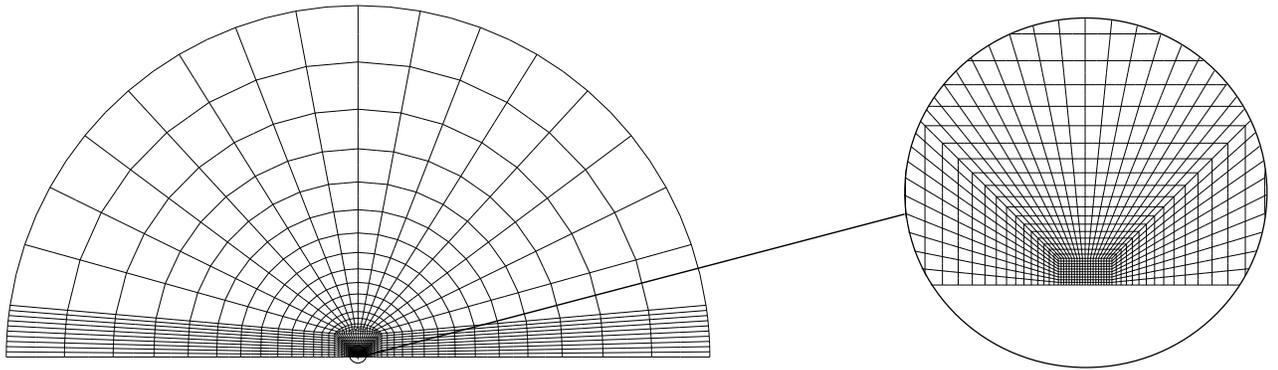}
\caption{Boundary layer formulation. Schematic and detailed view of the finite element mesh.}
\label{fig:Mesh}
\end{figure}

A remote $K$-field is prescribed by means of the so-called boundary layer formulation, see Fig. \ref{fig:Mesh}. Plane strain conditions are assumed. Consider both a polar coordinate system ($r$, $\theta$) and a Cartesian coordinate system ($x$, $y$) centred at the crack tip, with the crack plane along the negative $x$-axis. The outer $K$ field is imposed by prescribing the nodal displacements in the outer periphery of the mesh as,
\begin{equation}
u_x= \frac{1+\nu}{E} \sqrt{\frac{r}{2 \pi}} \left[ K_I  \left( 3 - 4 \nu - \cos \theta \right) \cos \left( \frac{\theta}{2} \right) + K_{II} \left( 5 - 4 \nu + \cos \theta \right) \sin \left( \frac{\theta}{2} \right) \right]
\end{equation}
\begin{equation}
u_y= \frac{1+\nu}{E} \sqrt{\frac{r}{2 \pi}}  \left[ K_I \left( 3 - 4 \nu - \cos \theta \right) \sin \left( \frac{\theta}{2} \right) + K_{II} \left( 1 - 4 \nu + \cos \theta \right) \cos \left( \frac{\theta}{2} \right)  \right]
\end{equation}

\noindent where $E$ is Young's modulus, $\nu$ is Poisson's ratio and $K_I$ and $K_{II}$ respectively denote the mode I and mode II stress intensity factors. Upon exploiting the symmetry about the crack plane, only half of the finite element model is analysed. After a mesh sensitivity study, the domain is discretised with 11,392 quadrilateral quadratic elements with full integration. As shown in Fig. \ref{fig:Mesh}, the mesh is progressively refined towards the crack tip to resolve the material strain gradient length $\ell$. From the outer $K$-field, a representative length of the plastic zone can be defined as,
\begin{equation}\label{eq:Irwin}
R_p=\frac{1}{3 \pi} \left( \frac{K}{\sigma_Y} \right)^2
\end{equation} 

We assume that the material obeys the following isotropic power-law hardening rule:
\begin{equation}
\sigma_F=\sigma_Y \left( 1 + \frac{E E^p}{\sigma_Y} \right)^N
\end{equation}

\noindent with the current flow stress, $\sigma_F$, being related to the gradient-enhanced effective plastic flow rate through the viscoplastic function - see (\ref{Eq:SfVEp}). The viscoplastic parameters are chosen to model the rate-independent limit. Specifically, following Ref. \cite{JMPS2020}, we define the following dimensionless constant:
\begin{equation}
c=\frac{\dot{K} \varepsilon_Y}{K \dot{\varepsilon}_0}
\end{equation}

\noindent where $\varepsilon_Y=\sigma_Y/E$ is the yield strain, and make suitable choices for $c$ and $m$. By comparing with the results obtained with rate-independent $J_2$ plasticity and the viscoplastic function by Panteghini and Bardella \cite{Panteghini2016} (with $\dot{\varepsilon}_0 \to 0$), we find that $c=0.25$ and $m=0.005$ accurately approximate the rate-independent limit. Throughout Section \ref{Sec:SSY}, material properties are assumed to be $\sigma_Y/E=0.003$, $\nu=0.3$ and $N=0.1$. We investigate the influence of $\chi$, the parameter that governs dissipation due to the plastic spin, and the ratio $\ell/R_p$, where $\ell$ is a reference length scale $L_E=L_D=\ell$. In addition, insight is gained into the role of the individual energetic $L_E$ and dissipative $L_D$ length scales.

\subsubsection{Asymptotic behaviour under Mode I fracture}

\begin{figure}[H]
\centering
\includegraphics[scale=1]{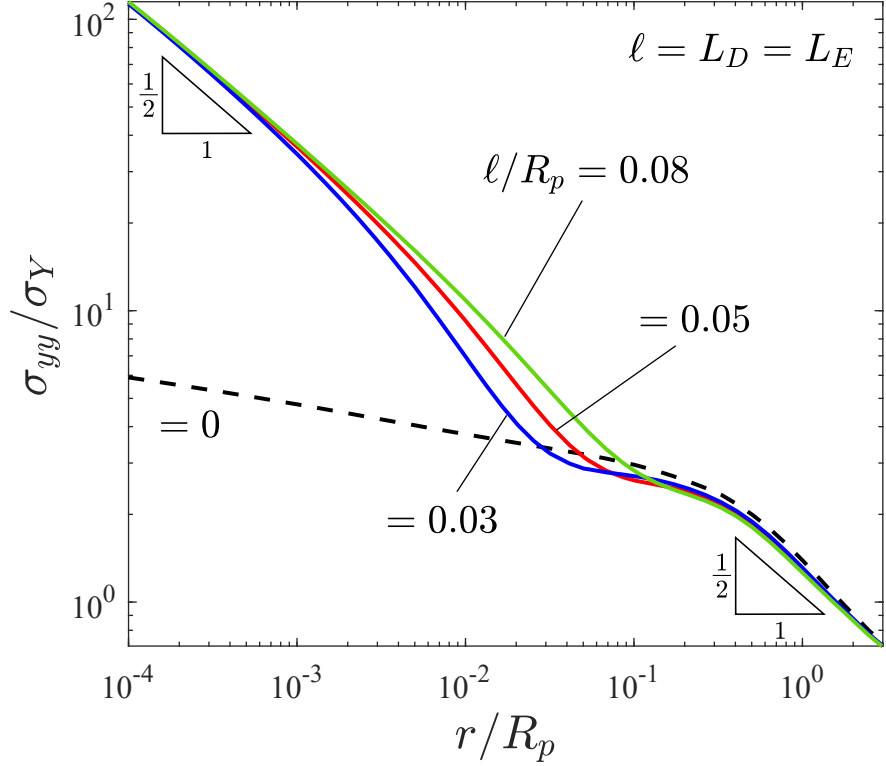}
\caption{Tensile stress distribution ahead of the crack tip for selected values of $\ell/R_p$. Material properties: $\sigma_Y/E=0.003$, $\nu=0.3$, $N=0.1$, and $\chi \to \infty$.}
\label{fig:StressLog}
\end{figure}

We proceed to verify the analytical findings of the asymptotic study in Section \ref{Sec:AResults}. Assume pure mode I conditions ($K_{II}=0$) and irrotational plastic flow ($\chi \to \infty$). The tensile stress distribution ahead of the crack tip is shown in log-log scale in Fig. \ref{fig:StressLog} for selected values of $\ell/R_p$. The finite element results confirm the analytical findings; for all $\ell/R_p>0$ values an elastic stress state exists close to the crack tip, where $\sigma_{yy}$ scales as $r^{-1/2}$. 

\begin{figure}[H]
\centering
\includegraphics[scale=1.05]{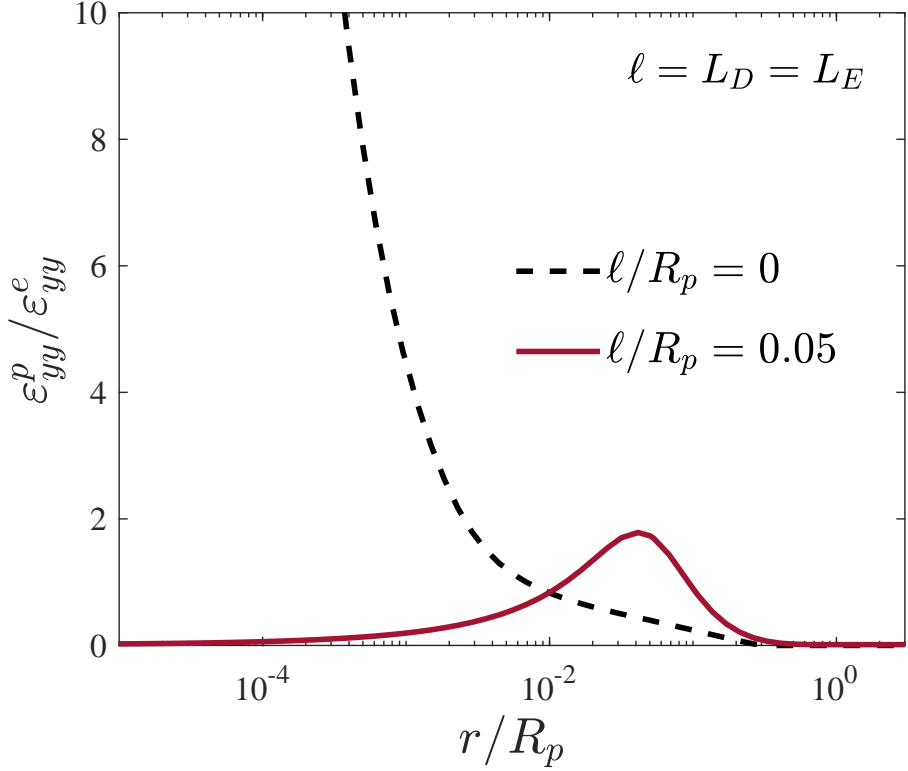}
\caption{Distribution ahead of the crack tip of the ratio between the tensile plastic and elastic strains $\varepsilon_{yy}^p/\varepsilon_{yy}^e$ for both conventional plasticity ($\ell/R_p=0$) and gradient plasticity ($\ell/R_p=0.05$). Material properties: $\sigma_Y/E=0.003$, $\nu=0.3$, $N=0.1$, and $\chi \to \infty$.}
\label{fig:Strain}
\end{figure}

Several distinct regions are seen in Fig. \ref{fig:StressLog}. Far away from the crack tip, the stress field is elastic and exhibits the linear elastic singularity $r^{-1/2}$. As the crack tip is approached a plastic region arises, where the stresses follow the HRR field of conventional $J_2$ plasticity \cite{Hutchinson1968,Rice1968}, with $\sigma_{yy}$ scaling as $r^{-N/(N+1)}$. This plastic region is reached at $r \approx 0.5 R_p$, as the Irwin approximation for the plastic zone length (\ref{eq:Irwin}) overestimates its size for strain hardening materials \cite{Anderson2005}. At approximately $r \leq \ell$, strain gradient hardening starts to play a role and a stress elevation is seen relative to the classic plasticity prediction (black dashed line, $\ell/R_p=0$). The size of the domain where gradient plasticity and conventional plasticity predictions deviate from each other is governed by $\ell/R_p$. Also, in the region $0.001R_p \leq r \leq 0.1 R_p$ the degree of stress elevation relative to conventional plasticity increases with $\ell/R_p$. However, in the vicinity of the crack tip ($r \approx 0.1 \ell$ or smaller), all $\ell/R_p>0$ cases superimpose, with the stress exhibiting the singular behaviour of linear elasticity $\sigma_{yy} \sim r^{-1/2}$, as predicted in the analytical asymptotic study. Note that path independence of the $J$-integral (\ref{eq:Jintegral}) implies that the outer and inner elastic $K$ fields must be the same; i.e., the inner $K$ field is identical for all $\ell/R_p>0$ values and corresponds to the one predicted by linear elasticity. Further insight into this elastic crack tip region is gained by plotting the ratio between the plastic strain and the elastic strain $\varepsilon_{yy}^p/\varepsilon_{yy}^e$; results are shown in Fig. \ref{fig:Strain} for conventional plasticity ($\ell/R_p=0$) and gradient plasticity ($\ell/R_p=0.05$).

As in Fig. \ref{fig:StressLog}, three regimes can be identified in Fig. \ref{fig:Strain}. For both $\ell=0$ and $\ell >0$, far away from the crack tip the plastic strains are zero but eventually increase as $r$ becomes smaller than $R_p$. In conventional plasticity the plastic strains are singular and raise sharply as we approach the crack tip. However, when $\ell>0$ the ratio $\varepsilon^p_{yy}/\varepsilon^e_{yy}$ reaches a peak and then drops, with the elastic strains dominating when $r \to 0$. An elastic strain (and stress) state exists near the crack tip, where plastic strains are negligible. Thus, the assumption of a dominating plastic strain field as $r \to 0$ cannot be used to derive the asymptotic fields, as done in the context of conventional plasticity. This crack tip elastic core resembles the concept of a dislocation-free zone \cite{Suo1993}.\\

\begin{figure}[H]
        \begin{subfigure}[h]{1.1\textwidth}
                \centering
                \includegraphics[scale=0.9]{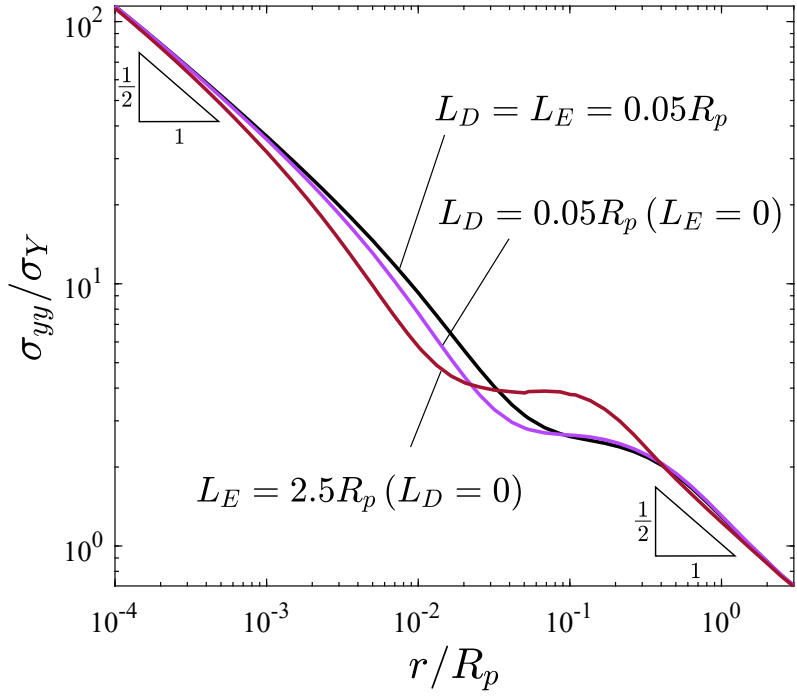}
                \caption{}
                \label{fig:StressLeLd}
        \end{subfigure}\\
		
        \begin{subfigure}[h]{1.1\textwidth}
                \centering
                \includegraphics[scale=0.75]{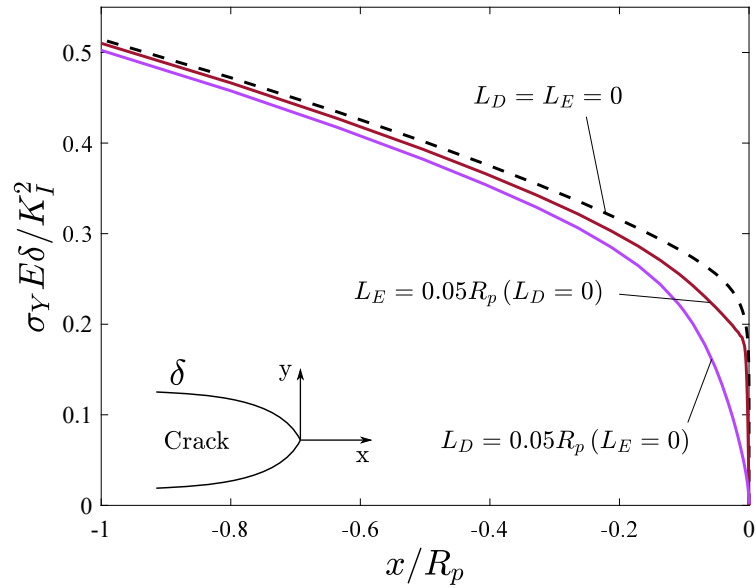}
                \caption{}
                \label{fig:CTODLeLd}
        \end{subfigure}       
        \caption{Influence of energetic and dissipative length scales: (a) tensile stress distribution ahead of the crack tip, and (b) crack opening profile. Material properties: $\sigma_Y/E=0.003$, $\nu=0.3$, $N=0.1$, and $\chi \to \infty$.}\label{fig:LevsLd}
\end{figure}

We proceed to assess the role of the individual energetic and dissipative higher order contributions. The crack tip stress distribution and the crack tip opening profile are respectively shown in Fig. \ref{fig:StressLeLd} and Fig. \ref{fig:CTODLeLd}. Dissipative higher order effects dominate the crack tip response; the magnitude of $L_E/R_p$ has to be increased 50 times relative to $L_D/R_p$ to achieve a similar degree of crack tip stress elevation. Given that we are under nearly proportional loading, differences must be due to the constitutive choices for the energetic defect stress (\ref{eq:defectstress}) and the dissipative higher order stress tensor (\ref{eq:DisspStress}c). In other words, the use of Nye's tensor as primal kinematic variable considerably reduces the local strengthening predicted ahead of a crack. The effect will likely be more profound if a less-than-quadratic defect energy is employed. Moreover, as it can be deduced from the analysis of Section \ref{Sec:AResults}, less-than-quadratic defect energies will change the nature of the singularity - the crack tip stress state will no longer be elastic if $L_D=0$. For the present formulation, where the defect energy is quadratic, the asymptotic behaviour described by the purely energetic result is the same as in the purely dissipative case; as shown analytically, if $L_E>0$ or $L_D>0$ the stress field exhibits the elastic singularity $r^{-1/2}$ as $r \to 0$. It is important to note that, in both the analytical and numerical analyses, the plastic distortion field is assumed to be continuous. However, for the case $L_D=0$ (where gradient effects are due to Nye's tensor only), the theoretical framework is characterised by kinematic higher order boundary conditions that admit discontinuity in some components of the plastic distortion. Thus, the results reported for the case $L_D=0$ should be taken with care; an $H$(curl) finite element framework, such as the one developed by Panteghini and Bardella \cite{Panteghini2018}, is needed to capture the discontinuities that might arise in Nye's tensor components. A very different outcome might be predicted if $\gamma_{ij}^p$ is allowed to be discontinuous and the solution localises.

\begin{figure}[H]
\centering
\includegraphics[scale=1]{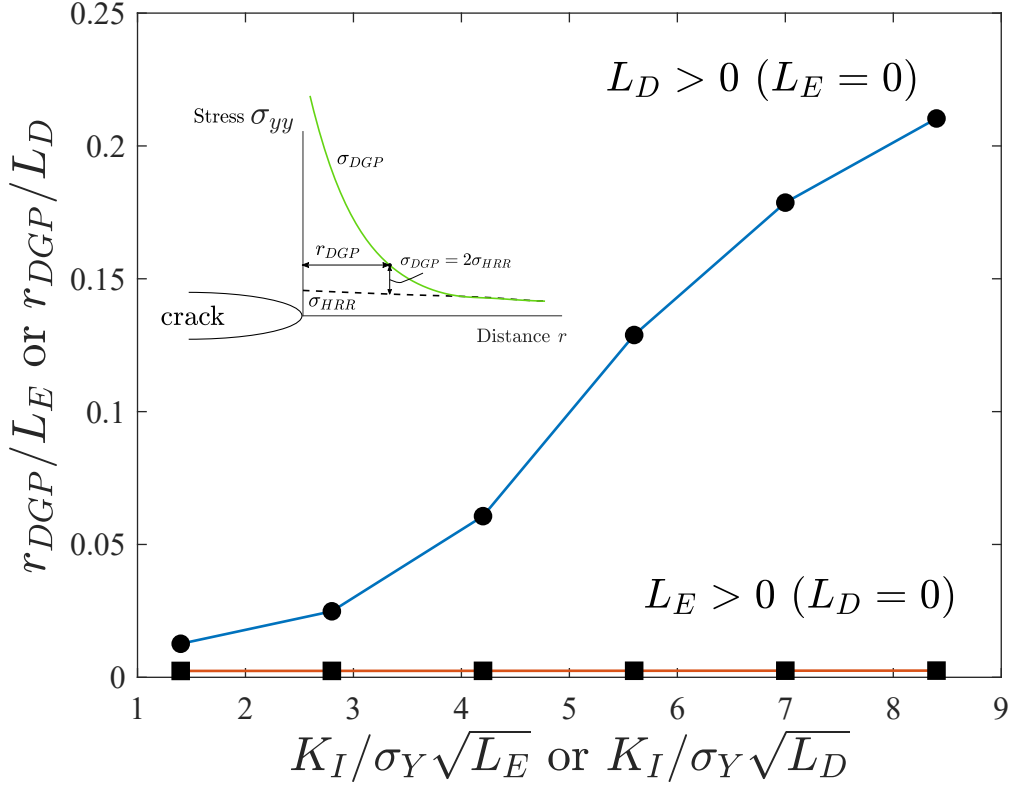}
\caption{Distance ahead of the crack tip where gradient effects significantly elevate the stresses relative to conventional plasticity ($\sigma_{DGP} > 2\sigma_{HRR}$). Material properties: $\sigma_Y/E=0.003$, $\nu=0.3$, $N=0.1$, and $\chi \to \infty$.}
\label{fig:Rdgp}
\end{figure}

As shown in Fig. \ref{fig:CTODLeLd}, the assumption of an equal magnitude for $L_E$ and $L_D$ leads to very different crack opening profiles. In the case of $L_D>0$ blunting is significantly reduced behind the crack tip. The crack profile also sharpens relative to the conventional plasticity prediction when $L_E>0$ ($L_D=0$) but to a much lesser extent. Far from the crack tip, the crack profile predictions for energetic gradient plasticity, dissipative gradient plasticity and conventional plasticity agree. Outside of the inner elastic core, the local strengthening predicted by Nye's tensor is much weaker than the one predicted by a gradient contribution based on the plastic strain gradient tensor. This is further explored in Fig. \ref{fig:Rdgp}, where the gradient dominated zone $r_{DGP}$ is plotted as a function of the remote mode I load $K_I$. As in Ref. \cite{IJP2016}, we define $r_{DGP}$ to represent the length of the region ahead of the crack tip where the stress distribution significantly deviates from conventional plasticity: $\sigma_{DGP} > 2\sigma_{HRR}$.

The results shown in Fig. \ref{fig:Rdgp} reveal a minor sensitivity of $r_{DGP}$ with the remote load for the case $L_E>0$ $(L_D=0)$. Tensile stresses are much higher than conventional plasticity in the elastic crack tip region but rapidly decay towards the conventional plasticity result farther away from the crack. Contrarily, in the case of $L_D >0$ the domain ahead of the crack where gradient effects significantly alter the stress distribution increases with the applied load. At the largest load level, the length of the stress elevation region is more than one order of magnitude larger if dissipative strengthening is accounted for. These differences are undoubtedly rooted in the choice of a free energy based on Nye's tensor. As shown in Ref. \cite{JMPS2019} for fracture and in Ref. \cite{Danas2012c} for bending, the dissipative contribution also outweighs the energetic counterpart when the defect energy is based on the plastic strain gradient tensor but differences are significantly smaller.\\

\begin{figure}[H]
\centering
\includegraphics[scale=1]{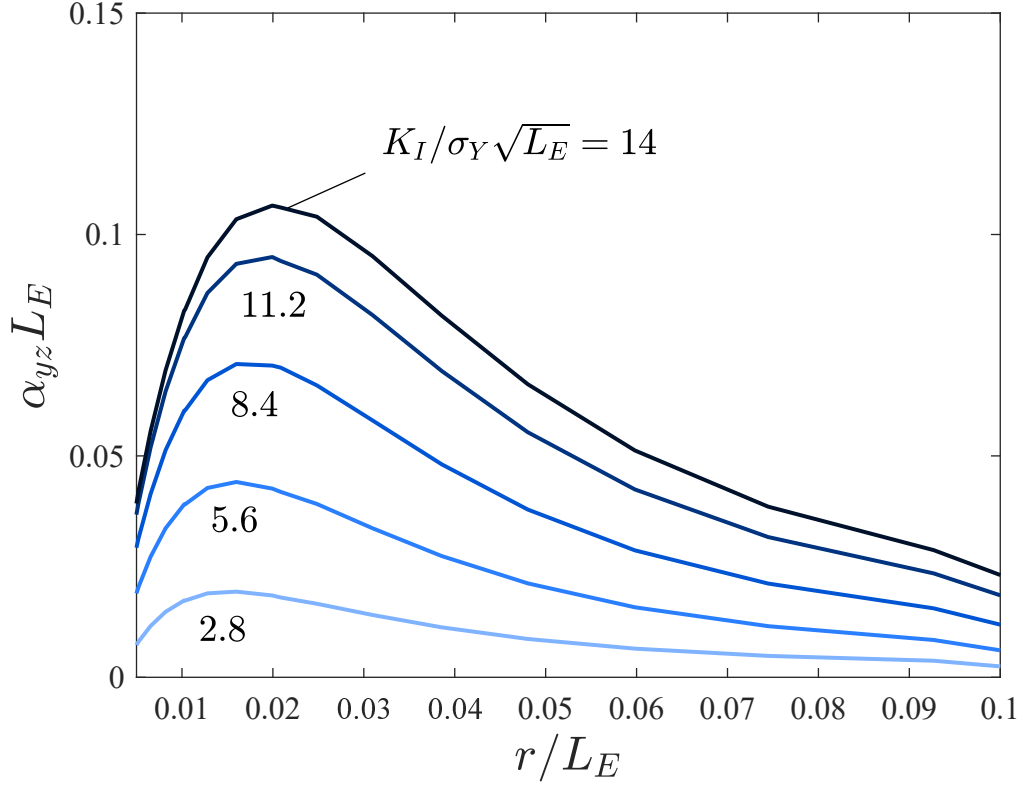}
\caption{Nye's tensor component $\alpha_{yz}$ distribution ahead of the crack tip for selected values of the remote load and the case $L_E>0$ ($L_D=0$). Material properties: $\sigma_Y/E=0.003$, $\nu=0.3$, $N=0.1$, and $\chi \to \infty$.}
\label{fig:Alpha23}
\end{figure}

Insight into the role of Nye's tensor is further gained by plotting the distribution ahead of the crack tip of the relevant component, $\alpha_{yz}$ - see Fig. \ref{fig:Alpha23}. The peak value of $\alpha_{yz}$ appears to saturate with an increasing remote load, reaching a maximum value on the order of $0.1/L_E$. Given that $L_E$ is typically within the 1-10 $\mu$m range (see Table \ref{tab:tab1length}), the maximum value of $\alpha_{yz}$ is on the order of 0.01-0.1 $\mu m^{-1}$, consistent with experimental observations \cite{Das2018}.  

The results shown in Figs. \ref{fig:LevsLd}-\ref{fig:Alpha23} reveal that, beyond the elastic inner core, a curl-based defect energy requires larger $L_E/R_p$ values to have a similar impact on the stress distribution as gradient-based constitutive choices. This raises the following question: what are the values of $L_E$ that fit the outcome of micro-scale experiments for curl- and gradient-based models? To the best of the authors' knowledge this data does not exist, motivating future work.\footnote{Bardella and Panteghini \cite{Bardella2015} obtained a good fit to the torsion experiments by Fleck \textit{et al.} \cite{Fleck1994} with a curl-based model but employed a logarithmic defect energy.}

\subsubsection{Mixed-model fracture - the role of $\chi$}

We proceed to assess the role of $\chi$ and the plastic spin. For that, mixed-mode fracture conditions are considered, where $K_I>0$ and $K_{II}>0$. The degree of mode-mixity can be characterised by the following angle:
\begin{equation}
\psi=\tan^{-1} \left( \frac{K_{II}}{K_I} \right)
\end{equation}

Note that, in the present model, the dissipation due to the plastic spin gradient is not accounted for. Accordingly, the plastic spin has no influence on the crack tip asymptotic behaviour; i.e., an elastic core exists, independently of the value of $\chi$. The plastic spin can play a role if we assume that the internal work is affected by different plastic rotations of two neighbouring macroscopic material points, as proposed by Bardella \cite{Bardella2010}. In the context of the original distortion gradient plasticity model \cite{Gurtin2004}, the influence of the plastic spin is limited to the stress elevation in the plastic region.\\

\begin{figure}[H]
\centering
\includegraphics[scale=1]{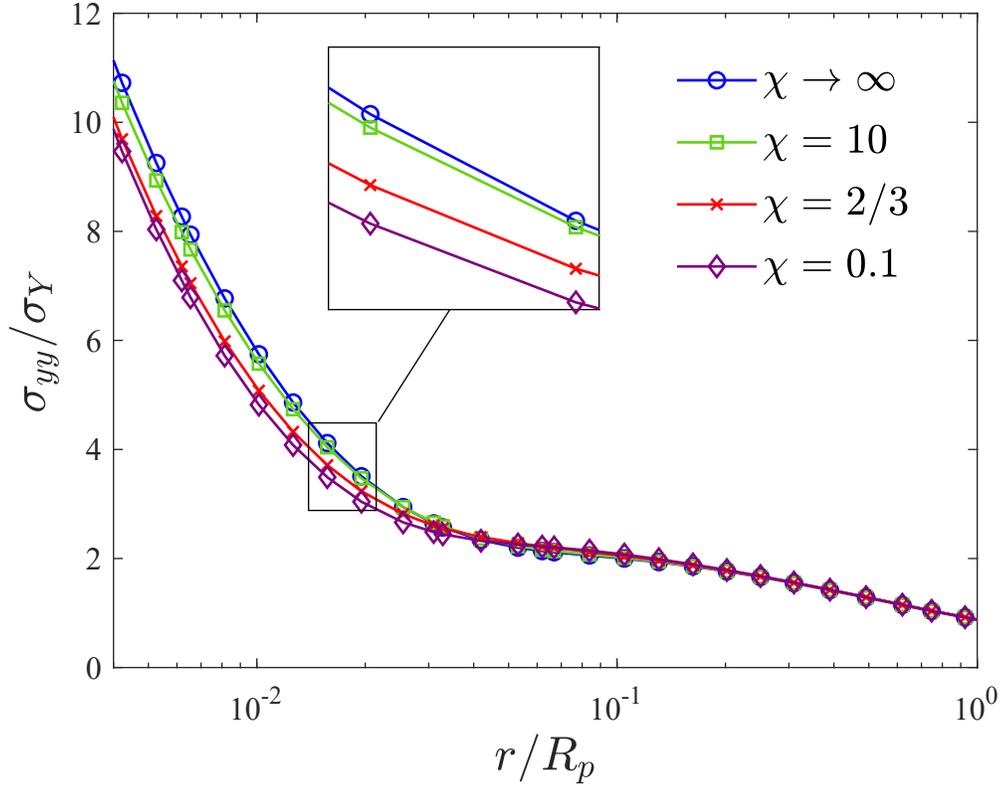}
\caption{Tensile stress distribution ahead of the crack tip under mode-mix conditions, $\psi=45^\circ$, for selected values of $\chi$. Material properties: $\sigma_Y/E=0.003$, $\nu=0.3$, $N=0.1$, and $\ell/R_p=0.03$.}
\label{fig:Chi}
\end{figure}

Crack tip stress fields are shown in Fig. \ref{fig:Chi} for $\psi=45^\circ$ and selected values of the parameter governing the dissipation due to the plastic spin, $\chi$. The choice $\chi=2/3$ makes the effective plastic flow rate (\ref{Eq:EpGurtin}) equal to the norm of the plastic distortion in the absence of higher order terms, while $\chi \to \infty$ reproduces the conditions of the theory by Gurtin and Anand \cite{Gurtin2005} (that is, irrotational plastic flow). The results reveal a small influence of the plastic spin, with the stress level increasing with $\chi$. This agrees with the trends observed by Bardella \cite{Bardella2010} in the simple shear problem, where augmenting $\chi$ leads to additional material hardening.

\begin{figure}[H]
\centering
\includegraphics[scale=0.9]{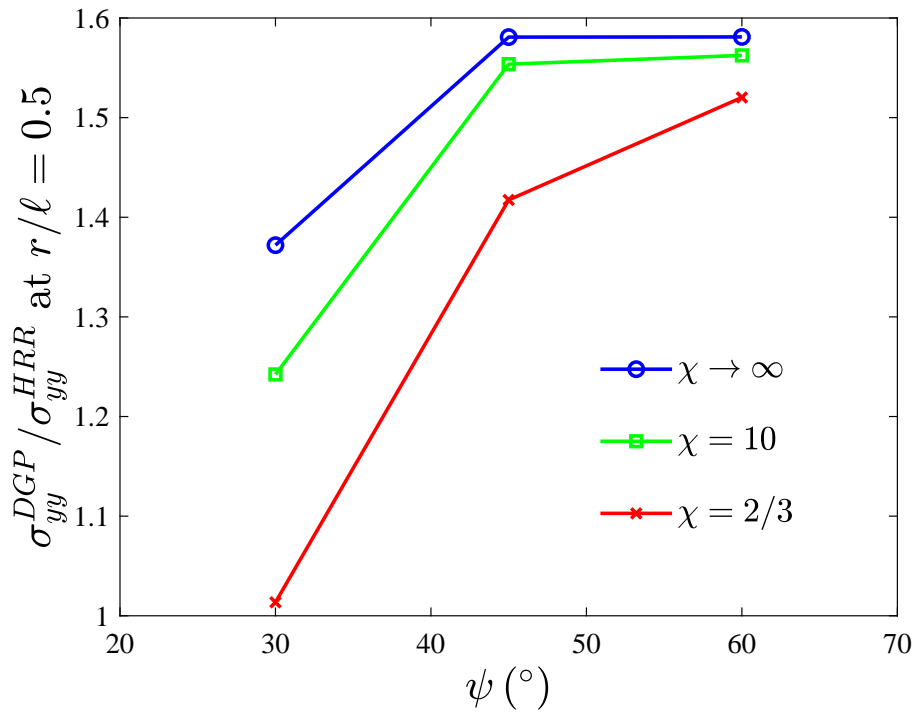}
\caption{Stress elevation predicted relative to conventional plasticity for selected values of $\chi$ and mode-mix angles. Material properties: $\sigma_Y/E=0.003$, $\nu=0.3$, $N=0.1$, and $\ell/R_p=0.03$.}
\label{fig:ChiElevation}
\end{figure}

The role of the plastic spin in elevating crack tip stresses is further investigated by computing the stress elevation relative to conventional plasticity $\sigma_{DGP}/\sigma_{HRR}$ as a function of the degree of mode mixity, as given by the angle $\psi$. The results are shown in Fig. \ref{fig:ChiElevation}. It is found that the influence of the plastic spin increases with decreasing $\psi$, and that the stress elevation increases with increasing $\psi$.

\subsection{Fracture at bi-material interfaces}
\label{Sec:Bimaterial}

We proceed to investigate cleavage in the presence of significant plastic flow at bi-material interfaces, a paradigmatic conundrum in metallic fracture. Specifically, we reproduce the classic experiments by Elssner \textit{et al.} \cite{Elssner1994} and Korn \textit{et al.} \cite{Korn2002} on niobium-sapphire interfaces. Our goal is to properly characterise the interface between elastic and elastic-plastic solids by incorporating, for the first time in fracture, the role of suitable higher order boundary conditions to model dislocation blockage.\\

\begin{figure}[H]
\centering
\includegraphics[scale=4.5]{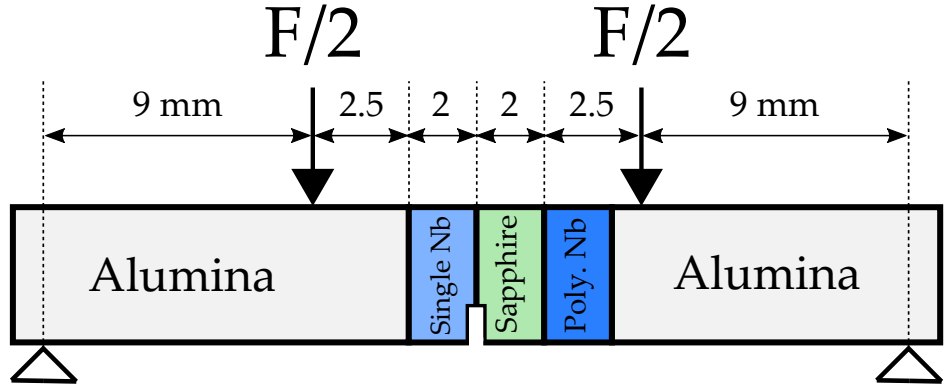}
\caption{Geometry and dimensions of the four-point bending experiments by Korn \textit{et al.} \cite{Korn2002}. All dimensions are in mm. The polycrystalline Nb layer has identical dimensions to the single crystal Nb and the sapphire layers. The specimen thickness (in the out-of-plane direction) is 2 mm.}
\label{fig:Bending}
\end{figure}

In a remarkable series of experiments, Elssner \textit{et al.} \cite{Elssner1994} and Korn \textit{et al.} \cite{Korn2002} measured both the macroscopic work of fracture and the atomic work of separation of an interface between sapphire and single crystal niobium. The macroscopic toughness turned out to be 1000 times higher than the atomic work of separation, with the difference being attributed to the significant dislocation activity observed in the Nb single crystal. However, fracture occurred by cleavage, with the crack tip remaining atomistically sharp. Since the stress level required to trigger atomic decohesion of a lattice or a strong interface is more than twice the maximum stress around the crack tip predicted by conventional $J_2$ plasticity, the findings by Elssner \textit{et al.} \cite{Elssner1994} and Korn \textit{et al.} \cite{Korn2002} constitute a paradox in the context of conventional plasticity \cite{Jiang2001,Qu2004,Jiang2010}. We hypothesize that strain gradient effects, combined with dislocation blockage, will be sufficient to raise crack tip stresses beyond the theoretical strength of the metal, $\approx 10 \sigma_Y$, over a sufficiently large distance to trigger fracture. The geometry, configuration and dimensions (in mm) of the four-point bending experiment Korn \textit{et al.} \cite{Korn2002} are shown in Fig. \ref{fig:Bending}. A sapphire layer is sandwiched between a single crystal and a polycrystalline Nb layers, which are in turn sandwiched by two alumina layers. An initial crack of length $a=0.4$ mm is placed at the interface between the single crystal Nb and sapphire. A load of $F=85$ N is applied.

Regarding the material properties, the alumina and the sapphire are linear elastic; alumina has a Young's modulus of $E=390$ GPa and a Poisson's ratio of $\nu=0.27$, while sapphire has a Young's modulus of $E=425$ GPa and a Poisson's ratio of $\nu=0.16$ \cite{Korn2002}. The polycrystalline niobium has a Young's modulus of $E=105$ GPa, a Poisson's ratio of $\nu=0.39$, a yield stress of $\sigma_Y=105$ MPa and a strain hardening exponent of $N=0.24$ \cite{Korn2002,Qu2004}. Furthermore, the single crystal niobium layer is characterised by a Young's modulus of $E=145$ GPa, a Poisson's ratio of $\nu=0.36$, and a yield stress of $\sigma_Y=145$ MPa; as in Ref. \cite{Qu2004}, the strain hardening exponent is chosen to be $N=0.05$ to represent easy glide in single crystal deformation. The magnitude of the material length scales remains to be defined. To the best of the authors' knowledge, no micro-scale experiments have been conducted on single crystal or polycrystalline Nb. A literature review of the experimental works conducted together with a gradient plasticity analysis to obtain the material length scales is provided in Table \ref{tab:tab1length}.

\begin{table}[H]
\caption[Literature review of experimentally reported length scales]{Compilation of experimentally reported length scales with their associated gradient plasticity formulation.}
\centering
\hspace*{-3cm} 
\begin{tabular}{c c c c} 
\hline
Work & Material & Experiment & Length scale - Gradient model\\
 \hline
Fleck \textit{et al.} \cite{Fleck1994} & Cu & Torsion & 3.7 $\mu m$ - Fleck \& Hutchinson (1993) \cite{Fleck1993} \\
Nix and Gao \cite{Nix1998} & Cu (cold worked) & Indentation & 5.84 $\mu m$ - Nix \& Gao (1998) \cite{Nix1998} \\
 & Single crystal Cu & Indentation & 12 $\mu m$ - Nix \& Gao (1998) \cite{Nix1998} \\
St\"{o}lken and Evans \cite{Stolken1998} & Ni & Bending & 5.2 $\mu m$ - Fleck \& Hutchinson (1993) \cite{Fleck1993} \\
Shrotriya \textit{et al.} \cite{Shrotriya2003} & Ni & Bending & 5.6 $\mu m$ - Fleck \& Hutchinson (1993) \cite{Fleck1993} \\
Haque and Saif \cite{Haque2003} & Al & Bending & 4.5 $\mu m$ - Gao \textit{et al.} (1999) \cite{Gao1999} \\
Ro \textit{et al.} \cite{Ro2006} & Al2024 & Indentation & 0.2 $\mu m$ - Nix \& Gao (1998) \cite{Nix1998} \\
Qian \textit{et al.} \cite{Qian2014} &  Steel S355 & Indentation & 7 $\mu m$ - Gao \textit{et al.} (1999) \cite{Gao1999}\\
 &  Steel S690 & Indentation & 7 $\mu m$ - Gao et al. (1999) \cite{Gao1999}\\
Guo \textit{et al.} \cite{Guo2017} &  Cu & Torsion & 3 $\mu m$ - Fleck \& Hutchinson (1993) \cite{Fleck1993}\\
Iliev \textit{et al.} \cite{Iliev2017} &  In & Indentation & 85.21 $\mu m$ - Nix \& Gao (1998) \cite{Nix1998}\\
 &  In & Bending & 93.34 $\mu m$ - Fleck \& Hutchinson (1993) \cite{Fleck1993} \\
Mu \textit{et al.} \cite{Mu2014} &  Cu & Micro-pillar shear & 0.647 $\mu m$ - Fleck \& Hutchinson (1997) \cite{Fleck1997}\\ 
 \hline
\end{tabular}
\label{tab:tab1length}
\end{table}
\pagestyle{plain}

\begin{figure}[H]
\centering
\includegraphics[scale=0.8]{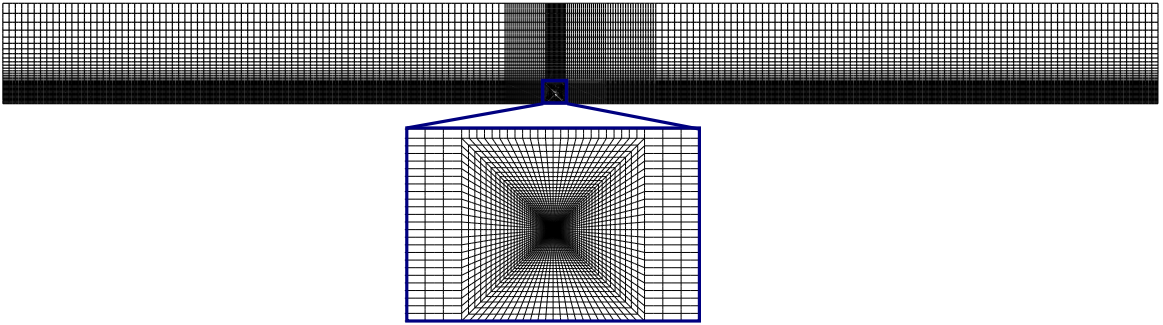}
\caption{Schematic and detailed views of the finite element mesh employed to model the four-point bending experiments by Korn \textit{et al.} \cite{Korn2002}.}
\label{fig:BendingMesh}
\end{figure}

As in Section \ref{Sec:SSY}, we consider a reference length scale $L_E=L_D=\ell$ for simplicity. We follow Qu \textit{et al.} \cite{Qu2004} and consider the magnitude of $\ell$ for polycrystalline Nb to be equal to 5.29 $\mu$m. This value is close to the average magnitude of the length scale measured for Cu, Ni and Al using torsion, indentation and bending. Regarding single crystal Nb, the work by Nix and Gao \cite{Nix1998} on Cu shows that experiments on single crystal samples are best captured with a length scale that duplicates the magnitude of the length scale employed to fit the tests on polycrystalline samples (see Table \ref{tab:tab1length}).  Accordingly, we chose to assume a length scale for single crystal Nb of $\ell=10.58$ $\mu$m. For both single crystal and polycrystalline Nb the parameter governing dissipation due to the plastic spin is assumed to be equal to $\chi=2/3$. The finite element model is constructed using user defined elements, for the elastic-plastic materials, and ABAQUS in-built elements for the elastic materials. A total of 20,336 quadrilateral quadratic elements with full integration are used, with the mesh being very refined close to the crack tip - see Fig. \ref{fig:BendingMesh}. The characteristic length of the elements close to the crack tip is of 10 nm. Plane strain conditions are assumed.

The higher order boundary conditions require special consideration. We assume that the  interaction between the dislocations and the various material interfaces in the bending specimen is such that dislocations are not allowed to exit the plastic layers.  These \textit{micro-hard} conditions, emulating dislocation blockage, are likely to be a good approximation to the dislocation behaviour at the niobium-sapphire and niobium-alumina interfaces. The degrees of freedom corresponding to the plastic strain tensor and the plastic spin are therefore constrained $\varepsilon^p_{x}=\varepsilon^p_y=\varepsilon_{xy}^p=\vartheta_{xy}^p=0$. Apart from that, the conventional boundary conditions are straightforward, as provided in Fig. \ref{fig:Bending}.\\

\begin{figure}[H]
\centering
\includegraphics[scale=1]{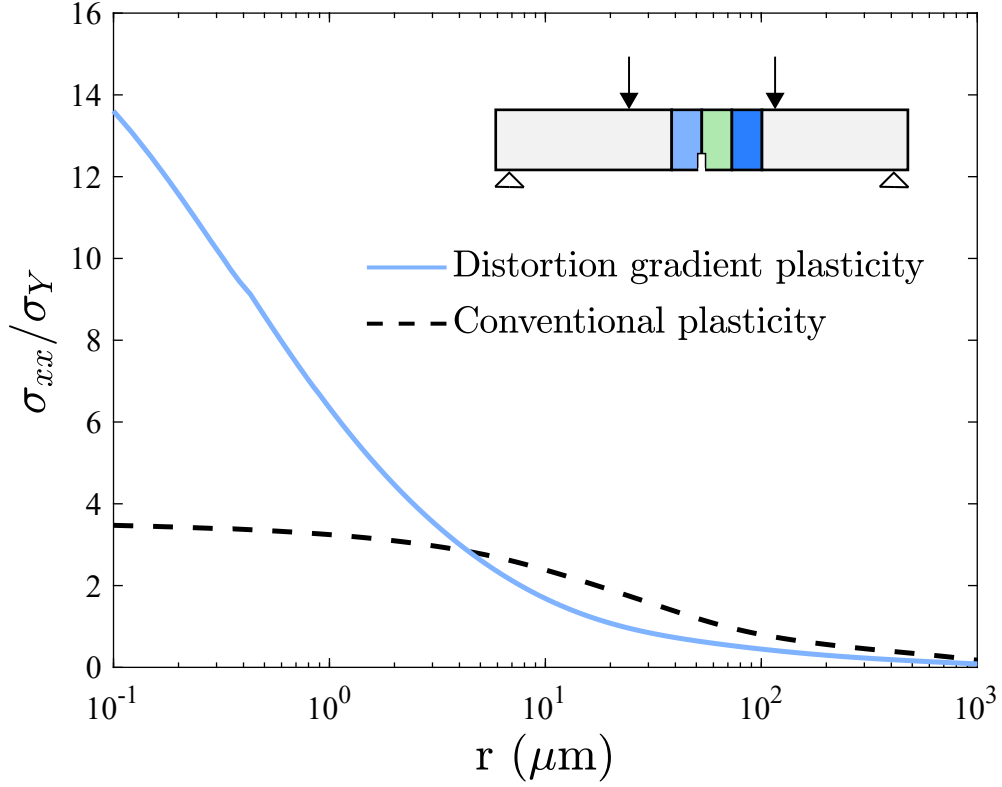}
\caption{Tensile stress distribution along the interface between single crystal Nb and sapphire ahead of the crack. The results are shown for distortion gradient plasticity (solid line) and for conventional plasticity (dashed line). The stress distribution is normalised by the yield stress of single crystal Nb.}
\label{fig:BendingStress}
\end{figure}

The results obtained from the finite element model are shown in Fig. \ref{fig:BendingStress}, in terms of tensile stress versus distance along the interface, ahead of the crack tip. The stress distribution is normalised by the yield stress of single crystal niobium and results are shown for both conventional and distortion gradient plasticity. While the maximum stress predicted by conventional plasticity is below 4$\sigma_Y$, insufficient to trigger brittle fracture, the stress level predicted with distortion gradient plasticity exceeds the theoretical lattice strength (10$\sigma_Y$) over hundreds of nanometres. Consequently, the combination of local crack tip strengthening and dislocation blockage provides a rational basis for atomic decohesion at bi-material interfaces in the presence of plasticity.

\begin{figure}[H]
\centering
\includegraphics[scale=0.5]{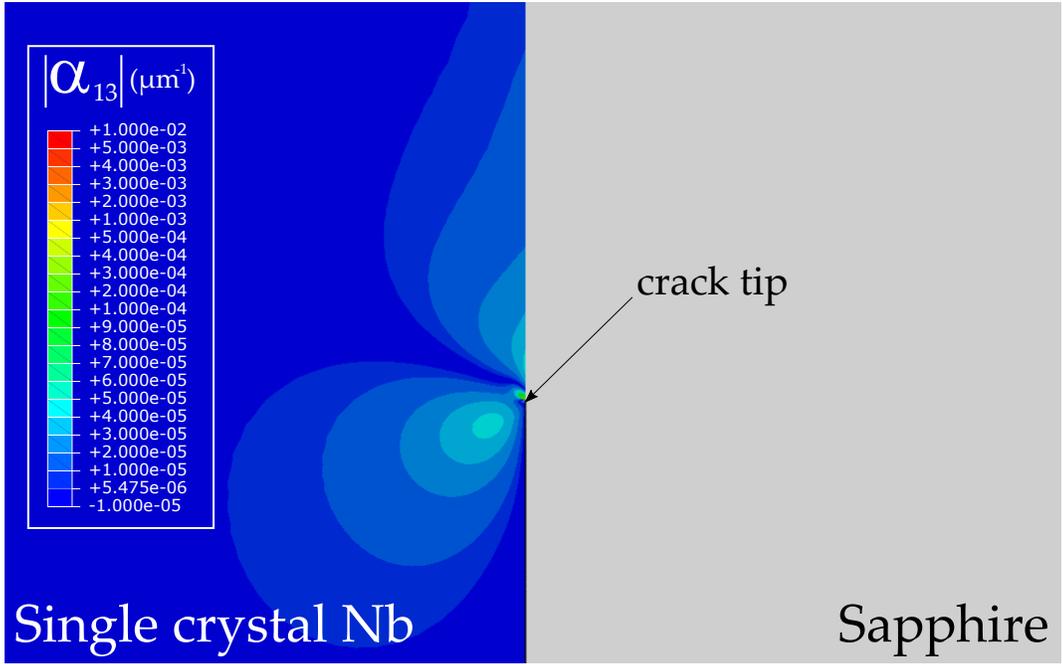}
\caption{Contours of the relevant component of Nye's tensor, $\alpha_{xz}$, in the Nb layer in the vicinity of the crack tip.}
\label{fig:NyeBending}
\end{figure}

Finally, contours are obtained for the relevant component of Nye's tensor, as shown in Fig. \ref{fig:NyeBending}. The magnitude of $\alpha_{xz}$ increases in the vicinity of the crack tip, reaching a maximum value that does not exceed 0.01 $\mu \textnormal{m}^{-1}$. This is in agreement with experimental observations of lattice distortions beneath nano-indents - see Ref. \cite{Das2018}.

\section{Conclusions}
\label{Sec:ConcludingRemarks}

We investigate, numerically and analytically, the crack tip behaviour of metals by using distortion gradient plasticity. The influence of two notable constitutive features on fracture mechanics predictions is investigated for the first time: (i) the use of Nye's tensor as primal higher order kinematic variable, and (ii) the role of the plastic spin. A generalised $J$-integral is defined, which is then used to determine the crack tip asymptotic singularity order. On the numerical side, a finite element framework is presented, which builds upon a novel viscoplastic function that enables efficient modelling of both rate-dependent and rate-independent behaviour. The analysis of crack tip fields under mode I and mixed-mode fracture assuming small scale yielding conditions reveals the following main findings:

\begin{itemize}

\item An \emph{elastic} region exists close to the crack tip, where the plastic strains are negligible and the Cauchy stress follows the $r^{-1/2}$ singularity of linear elasticity. This elastic core, reminiscent of a dislocation-free zone, is present for both gradient-based and curl-based higher order constitutive choices. However, the nature of the singularity will change for a less-than-quadratic defect energy.

\item The stress elevation due to strain gradient hardening predicted in the annular plastic zone embedding the elastic core is very sensitive to the constitutive choice of the defect energy. The use of Nye's tensor leads to a substantially weaker stress elevation, as compared with the plastic strain gradient tensor, for the same value of the material length scale. The values of Nye's tensor predicted in the vicinity of the crack are consistent with experimental observations.

\item A small influence of $\chi$, the parameter governing the dissipation due to the plastic spin, is observed. Increasing $\chi$ raises crack tip stresses, with the upper bound being given by the irrotational plastic flow scenario.

\end{itemize}

We emphasise that, for the case where gradients effects are due to Nye's tensor only ($L_D=0$), our analysis is constrained by the assumption of a continuous plastic distortion field. The framework should be extended to admit discontinuity in some components of the plastic distortion, as in \cite{Panteghini2018}, and this will be the goal of future work.

Finally, the framework is employed to shed light into the paradox of brittle fracture in the presence of plasticity in bi-material interfaces. By modelling the paradigmatic experiments by Elssner \textit{et al.} \cite{Elssner1994} and Korn \textit{et al.} \cite{Korn2002} on niobium-sapphire interfaces, we find that:

\begin{itemize}

\item The combination of micro-hard higher order boundary conditions, emulating dislocation blockage, and gradient plasticity effects lead to interface crack tip stresses that are larger than the theoretical lattice strength over a distance of hundreds of nm; rationalising quasi-cleavage in bi-material interfaces.

\end{itemize}

\section{Acknowledgments}
\label{Sec:Acknowledgeoffunding}

Helpful discussions with Konstantinos Poulios (Technical University of Denmark) and Ivan Moyano (University of Cambridge) are gratefully acknowledged. E. Mart\'{\i}nez-Pa\~neda acknowledges financial support from the People Programme (Marie Curie Actions) of the European Union's Seventh Framework Programme (FP7/2007-2013) under REA grant agreement n$^{\circ}$ 609405 (COFUNDPostdocDTU).

\appendix

\section{Additional details of numerical implementation}
\label{App:FEM}

Assume 2D plane strain conditions, as in the numerical examples addressed in the paper. Accordingly, for an element with $k$ nodes, the nodal variables read,
\begin{equation}
\boldsymbol{\hat{u}}= \begin{bmatrix} \hat{u}^{(1)}_x & \hat{u}^{(1)}_y & \cdots & \hat{u}^{(k)}_x & \hat{u}^{(k)}_y \end{bmatrix}^T
\end{equation}
\begin{equation}
\boldsymbol{\hat{\varepsilon}}^p= \begin{bmatrix} \hat{\varepsilon}^{p \, (1)}_x & \hat{\varepsilon}^{p \, (1)}_y & \hat{\gamma}^{p \, (1)}_{xy} & \cdots & \hat{\varepsilon}^{p \, (k)}_x & \hat{\varepsilon}^{p \, (k)}_y & \hat{\gamma}^{p \, (k)}_{xy} \end{bmatrix}^T
\end{equation}
\begin{equation}
\hat{\vartheta}^p_{xy}= \begin{bmatrix} \hat{\vartheta}^{p \, (1)}_{xy} &  \cdots & \hat{\vartheta}^{p \, (k)}_{xy} \end{bmatrix}^T
\end{equation}

The shape functions matrices for a given node $n$ are then given by,
\begin{equation}
\boldsymbol{N}_n^{\bm{u}} = \begin{bmatrix} N_n & 0 \\
0 & N_n \end{bmatrix}; \,\,\,\,\,\,\,\,\,\,\,\,\,\,\,\, 
\boldsymbol{N}_n^{\bm{\varepsilon}^p}= \begin{bmatrix}
N_n & 0 & 0  \\
0 & N_n & 0 \\
-N_n & -N_n & 0  \\
0 & 0 & N_n 
\end{bmatrix}
\end{equation}

\noindent with $\boldsymbol{N}_n^{\bm{\vartheta}^p}$ being, in plane strain conditions, the scalar $N_n$ for node $n$. While the interpolation matrices for gradient and curl-based quantities are given by,
\begin{equation}
\boldsymbol{B}^{\bm{u}}_n=\begin{bmatrix} \frac{\partial N_n}{\partial x} & 0 \\
0 & \frac{\partial N_n}{\partial y} \\
0 & 0 \\
\frac{\partial N_n}{\partial y} & \frac{\partial N_n}{\partial x} 
\end{bmatrix}; \,\,\,\,\,\,\,\,\,\,\,\,\,\,\,\,\,\,
 \boldsymbol{B}^{\bm{\varepsilon}^p}_n=\begin{bmatrix}
\frac{\partial N_n}{\partial x} & 0 & 0 \\
\frac{\partial N_n}{\partial y} & 0 & 0  \\
0 & \frac{\partial N_n}{\partial x} & 0  \\
0 & \frac{\partial N_n}{\partial y} & 0  \\
-\frac{\partial N_n}{\partial x} & -\frac{\partial N_n}{\partial x} & 0 \\
-\frac{\partial N_n}{\partial y} & -\frac{\partial N_n}{\partial y} & 0 \\
0 & 0 & \frac{\partial N_n}{\partial x}  \\
0 & 0 & \frac{\partial N_n}{\partial y} 
\end{bmatrix};
\end{equation}
\noindent and,
\begin{equation}
\boldsymbol{M}^{\bm{\varepsilon}^p}_n=\begin{bmatrix}
-\frac{\partial N_n}{\partial y} & 0 & \frac{1}{2} \frac{\partial N_n}{\partial x} \\
0 & \frac{\partial N_n}{\partial x} & -\frac{1}{2} \frac{\partial N_n}{\partial y}  \\
-\frac{\partial N_n}{\partial y}  & -\frac{\partial N_n}{\partial y}  & 0  \\
\frac{\partial N_n}{\partial x} & \frac{\partial N_n}{\partial x} & 0 
\end{bmatrix};  \,\,\,\,\,\,\,\,\,\,\,\,\,\,\,\,\,\,
\boldsymbol{M}^{\bm{\vartheta}^p}_n=\begin{bmatrix}
\frac{\partial N_n}{\partial x} \\
\frac{\partial N_n}{\partial y}  \\
 0  \\
 0 
\end{bmatrix}
\end{equation}\\

On the other side, the stiffness matrix components are given by,
\begin{equation}
\bm{K}^{u,u}_{nm} =\frac{\partial \bm{R}^u_n}{\partial \bm{u}_m} = \int_{\Omega}  \left( \boldsymbol{B}_n^{\bm{u}} \right)^T \bm{C} \, \boldsymbol{B}_m^{\bm{u}} \, \textnormal{d}V
\end{equation}
\begin{equation}
\bm{K}^{u,\varepsilon^p}_{nm} =\frac{\partial \bm{R}^u_n}{\partial \bm{\varepsilon}^p_m} = - \int_{\Omega} \left( \boldsymbol{B}_n^{\bm{u}} \right)^T \bm{C} \boldsymbol{N}_m^{\bm{\varepsilon}^p} \, \textnormal{d}V
\end{equation}
\begin{equation}
\bm{K}^{\varepsilon^p, u}_{nm} =\frac{\partial \bm{R}^{\bm{\varepsilon}^p}_n}{\partial \bm{u}_m} = - \int_{\Omega} \left( \boldsymbol{N}_n^{\bm{\varepsilon}^p} \right)^T \bm{C}  \, \boldsymbol{B}_m^{\bm{u}} \, \textnormal{d}V
\end{equation}
\begin{align}\label{Eq:Kepep} 
\bm{K}^{\varepsilon^p, \varepsilon^p}_{nm} =  &\frac{\partial \bm{R}^{\bm{\varepsilon}^p}_n}{\partial \bm{\varepsilon}^p_m} = \int_{\Omega} \Bigg\{ \left( \boldsymbol{N}_n^{\bm{\varepsilon}^p} \right)^T \left[ \left( \frac{\partial \bm{q}}{\partial \varepsilon^p_m} + \bm{C} \right) \boldsymbol{N}_m^{\bm{\varepsilon}^p} + \frac{\partial \bm{q}}{\partial \nabla \varepsilon^p_m}  \boldsymbol{B}_m^{\bm{\varepsilon}^p} \right] \nonumber \\
& + \left(  \boldsymbol{B}_n^{\bm{\varepsilon}^p} \right)^T \left( \frac{\partial \bm{\tau}}{\partial \varepsilon^p_m} \boldsymbol{N}_m^{\bm{\varepsilon}^p} + \frac{\partial \bm{\tau}}{\partial \nabla \varepsilon^p_m} \boldsymbol{B}_m^{\bm{\varepsilon}^p} \right) + \left( \boldsymbol{M}_n^{\bm{\varepsilon}^p} \right)^T \frac{\partial \bm{\zeta}}{\partial \alpha_m} \boldsymbol{M}_m^{\bm{\varepsilon}^p} \Bigg\} \, \textnormal{d}V
\end{align}
\begin{equation}
\bm{K}^{\varepsilon^p, \vartheta^p}_{nm} =\frac{\partial \bm{R}^{\bm{\varepsilon}^p}_n}{\partial \bm{\vartheta}^p_m} = \int_{\Omega} \left[ \left( \boldsymbol{N}_n^{\bm{\varepsilon}^p} \right)^T \frac{\partial \bm{q}}{\partial \vartheta_m^p} \boldsymbol{N}_m^{\bm{\vartheta}^p} + \left( \boldsymbol{B}_n^{\bm{\varepsilon}^p} \right)^T \frac{\bm{\partial \tau}}{\partial \vartheta_m^p} \boldsymbol{N}_m^{\bm{\vartheta}^p} + \left( \boldsymbol{M}_n^{\bm{\varepsilon}^p} \right)^T  \frac{\partial \zeta}{\partial \alpha_m} \boldsymbol{M}_m^{\bm{\vartheta}^p} \right] \, \textnormal{d}V
\end{equation}
\begin{equation}
\bm{K}^{\vartheta^p, \varepsilon^p}_{nm} =\frac{\partial \bm{R}^{\bm{\vartheta}^p}_n}{\partial \bm{\varepsilon}^p_m} = \int_{\Omega} \left[ \left( \boldsymbol{N}_n^{\bm{\vartheta}^p} \right)^T \left( \frac{\partial \bm{\omega}}{\partial \varepsilon_m^p} \boldsymbol{N}_m^{\bm{\varepsilon}^p} + \frac{\partial \bm{\omega}}{\partial \nabla \varepsilon^p_m} \boldsymbol{B}_m^{\bm{\varepsilon}^p} \right)  + \left( \boldsymbol{M}_n^{\bm{\vartheta}^p} \right)^T  \frac{\partial \zeta}{\partial \alpha_m} \boldsymbol{M}_m^{\bm{\varepsilon}^p} \right] \, \textnormal{d}V
\end{equation}
\begin{equation}
\bm{K}^{\vartheta^p, \vartheta^p}_{nm} =\frac{\partial \bm{R}^{\bm{\vartheta}^p}_n}{\partial \bm{\vartheta}^p_m} = \int_{\Omega} \left[ \left( \boldsymbol{N}_n^{\bm{\vartheta}^p} \right)^T  \frac{\partial \bm{\omega}}{\partial \vartheta_m^p} \boldsymbol{N}_m^{\bm{\vartheta}^p}  + \left( \boldsymbol{M}_n^{\bm{\vartheta}^p} \right)^T  \frac{\partial \zeta}{\partial \alpha_m} \boldsymbol{M}_m^{\bm{\vartheta}^p} \right] \, \textnormal{d}V
\end{equation}



\bibliographystyle{elsarticle-num}
\bibliography{library}

\end{document}